\PassOptionsToPackage{x11names}{xcolor}
\documentclass[12pt,a4paper,twoside]{article}
\usepackage[colorinlistoftodos, textwidth=30mm, shadow]{todonotes}

\usepackage{array}
\usepackage{amsmath}
\usepackage{tabularx}
\usepackage{txfonts}
\usepackage{enumerate}
\usepackage{float}
\usepackage{xspace}
\usepackage[x11names]{xcolor}
\usepackage{multirow}
\usepackage{algpseudocode,algorithmicx}
\usepackage{booktabs}
\usepackage{siunitx}

\usepackage{subcaption}

\usepackage{pdflscape}

\usepackage[T1]{fontenc}
\usepackage[cp1250]{inputenc}

\usepackage[margin=2.5cm]{geometry}

\usepackage[colorlinks=true,citecolor=red,linkcolor=DarkOrange1]{hyperref}

\usepackage{changes}

\usepackage{listings}
\usepackage{placeins} 

\newtheorem{theorem}{Theorem}[section]
\newtheorem{lemma}[theorem]{Lemma}

\numberwithin{equation}{section}

\begin{document}
\pagestyle{myheadings}
\markright{M.J.~Piotrowska et al.}

\noindent\begin{tabular}{|p{\textwidth}}
	\Large\bf  
	Modelling pathogen spread in a healthcare network: indirect patient movements
 \\\vspace{0.01cm}
 
    \it Piotrowska, M.J.$^{\dagger,\ddagger,1}$, Sakowski, K.$^{\dagger,*,\diamondsuit,\ddagger,2}$, Karch, A.$^{\star,3}$, Tahir, H.$^{\clubsuit,4}$,  Horn, J.$^{\natural,5}$, Kretzschmar, M.E.$^{\clubsuit,\heartsuit,6}$, Mikolajczyk, R.T.$^{\natural,7}$ 
    \\\vspace{0.02cm}
\it\small $^\dagger$Institute of Applied Mathematics and Mechanics,\\
\it\small University of Warsaw, Banacha 2, 02-097 Warsaw, Poland\\\vspace{0.01cm}
\it\small $^*$Institute of High Pressure Physics,\\
\it\small Polish Academy of Sciences, Sokolowska 29/37, 01-142 Warsaw, Poland\\\vspace{0.01cm}
\it\small $^\diamondsuit$Research Institute for Applied Mechanics,\\
\it\small Kyushu University, Kasuga, Fukuoka 816-8580, Japan\\\vspace{0.01cm}
\it\small $^\star$ Institute for Epidemiology and Social Medicine,\\ 
\it\small Albert-Schweitzer-Campus 1, Geb{\"a}ude D3, 48149 M\"{u}nster, Germany \\\vspace{0.01cm}
\it\small $\natural$ Medical Epidemiology, Biometrics and Informatics,\\
\it\small Martin Luther University Halle-Wittenberg, Magdeburger Strasse 8,
06112 Halle (Saale), Germany\\\vspace{0.01cm}
\it\small $^\clubsuit$ Julius Center for Health Sciences \& Primary Care, University Medical Center Utrecht,\\ Utrecht University,
\it\small Heidelberglaan 100, 3584 CX Utrecht, The Netherlands \\\vspace{0.01cm}
\it\small $^\heartsuit$ Centre for Infectious Disease Control, National Institute of Public Health and the Environment (RIVM),\\
\it\small Antonie van Leeuwenhoeklaan 9, 3721 MA Bilthoven, The Netherlands \\\vspace{0.01cm}

\small  $^1$\texttt{monika@mimuw.edu.pl}, $^2$\texttt{konrad@mimuw.edu.pl}, $^3$\texttt{Andre.Karch@ukmuenster.de}, $^4$\texttt{h.tahir@umcutrecht.nl},
$^5$\texttt{johannes.horn@uk-halle.de},
$^6$\texttt{m.e.e.kretzschmar@umcutrecht.nl},
$^7$\texttt{rafael.mikolajczyk@uk-halle.de}\\
\small $\ddagger$ equal contribution\\
	\hline
\end{tabular}
\thispagestyle{empty}

\begin{abstract}
	A hybrid network--deterministic model for simulation of multiresistant pathogen spread in a healthcare system is presented. The model accounts for two paths of pathogen transmission between the healthcare facilities: inter-hospital patient transfers (direct transfers) and readmission of colonized patients (indirect transfers). In the latter case, the patients may be readmitted to the same facility or to a different one. Intra-hospital pathogen transmission is governed by a SIS model expressed by a system of ordinary differential equations.

	Using a network model created for a Lower Saxony region (Germany), we showed that the proposed model reproduces the basic properties of healthcare-associated pathogen spread.
	Moreover, it shows the important contribution of the readmission of colonized patients on the prevalence of individual hospitals as well as of overall healthcare system: it can increase the overall prevalence by the factor of 4 as compared to inter-hospital transfers only.
	The final prevalence in individual healthcare facilities was shown to depend on average length of stay by a non-linear concave function.

	Finally, we demonstrated that the network parameters of the model may be derived from administrative admission/discharge records. In particular, they are sufficient to obtain inter-hospital transfer probabilities, and to express the patients' transfer as a Markov process.
\end{abstract}


\tableofcontents
\addtocontents{toc}{\protect\setcounter{tocdepth}{3}}
\noindent\begin{tabular}{p{\textwidth}}
	\\
	\hline
\end{tabular}
\vspace{2em}\\

\section{Introduction}
	Recent years brought an increased attention to the question of how the patient traffic between healthcare facilities contributes to the spread of healthcare-associated infections in general and multidrug resistant pathogens in particular~\cite{Ciccolini2014a}. The movements of patients between hospitals can be divided into transfers of patients from one hospital to another (i.e. direct transfers), and in readmission of patients to the same or another hospital after having spent some time in the community (i.e. indirect transfers). Both direct and indirect transfers may contribute to the spread of pathogens in healthcare networks. Interventions to prevent the spread of pathogens in a network may differ for direct and indirect transfers. While screening of patients, who are transferred from one hospital to another is an obvious intervention measure, this is less clear for patients who are readmitted from the community~\cite{Roth2015}. The effectiveness/cost-effectivenes of screening of patients after an indirect transfer depends on the time between admissions, the clearance rate of the pathogen, and whether the screening is applied in targeted way, i.e. based on individual risk factors of the patient. To quantify the effectiveness of such measures, we need to understand the contribution of indirect transfers to the spread of pathogens through the network. In the past decade, eleven studies assessed healthcare networks in countries or federal states. All of these used national or federal registries in the US \cite{Karkada2011} \cite{Lee2011} \cite{Simmering2015} \cite{Fernandez-Gracia2017} \cite{Ray2018} or Europe (England, France, Germany, the Netherlands) \cite{Donker2010} \cite{Donker2012} \cite{Ciccolini2014} \cite{Belik2016} \cite{Nekkab2017} \cite{Donker2019}. The definition of transfer and patient movement was quite heterogeneous. While some studies took into account direct and indirect transfers independently of the length of a community stay between two hospital visits \cite{Donker2010} \cite{Belik2016}, others restricted indirect transfers to a maximum of 90 \cite{Ray2018} or 365 days~\cite{Donker2019}, or did not consider them at all \cite{Fernandez-Gracia2017} when deriving hospital network flow characteristics. This heterogeneity makes comparisons of the findings difficult and also sometimes the description of how the indirect transfers were incorporated is not clear. For Germany, only preliminary analyses were conducted \cite{Belik2016}.
	Using data from one regional health care insurance in Germany, we develop a model explicitely accounting for indirect transfers and describe selected implications of either including or ignoring indirect transfers for the spread of mutliresistant pathogens in the health care network, using the example of methicillin-resistant Staphylococcus aureus (MRSA), the best studied healthcare-associated pathogen.

\section{Materials and Methods}\label{sec:matandmeth}

	\subsection{Description of dataset}\label{sec:description:of:data:set}

		In German health care system over 90\% of patients are insured in public insurance companies. The remaining few percent with top income have private insurances. Overall, more than 190 public insurance companies exist. This scattered data was a reason that reimbursement data were less accessible for scientific analysis than in other countries, in addition to high standards of data protection, which limited their use. 
		We used an anonymized hospital discharge database provided by AOK Lower Saxony, a statuary regional healthcare insurance company in Germany. AOK Lower Saxony includes almost exclusively persons living in the federal state of Lower Saxony and covers around 30\% of the local population.
	
		The dataset used contains hospitalisation records of 1 673 247 patients for the years 2008 to 2015. For each hospital stay the anonymized patient ID, the anonymized healthcare facility ID, the federal state where the healthcare facility is located, day of admission, day of discharge, discharge diagnosis (ICD 10 GM code) as well as age and sex of the patient are available. For data protection reasons, information on exact geographical location of the healthcare facilities is not contained in the provided database. In the dataset, we identified 4 573 584 hospitalisations in 223 facilities located in Lower Saxony, and 680 908 for healthcare facilities in other German federal states.
		
		Eight of these healthcare facilities were excluded from all further analyses because of insufficient data (some of them had very small admission number over the period of eight years,  or there were not in operation (no patients registered in the input data) for significant time interval).  Thus, we finally consider 164 healthcare facilities located in Lower Saxony. Moreover, records without discharge diagnosis code were omitted in our analysis (about 6\% of all records). Among them, there were 257\,668 hospitalisation records for facilities in Lower Saxony and 79\,102 records for facilities in the remaining parts of Germany.

		For more detailed information on e.g. sex of patients, distributions of length of patient stays, detailed characteristics of overlaps as well as a technical description of the provided database, we refer to a technical report~\cite{Piotrowska2019}.

	\subsection{Model input data} \label{sec:data:analysis}

\begin{figure}
			\centering
			\includegraphics[width=0.5\linewidth]{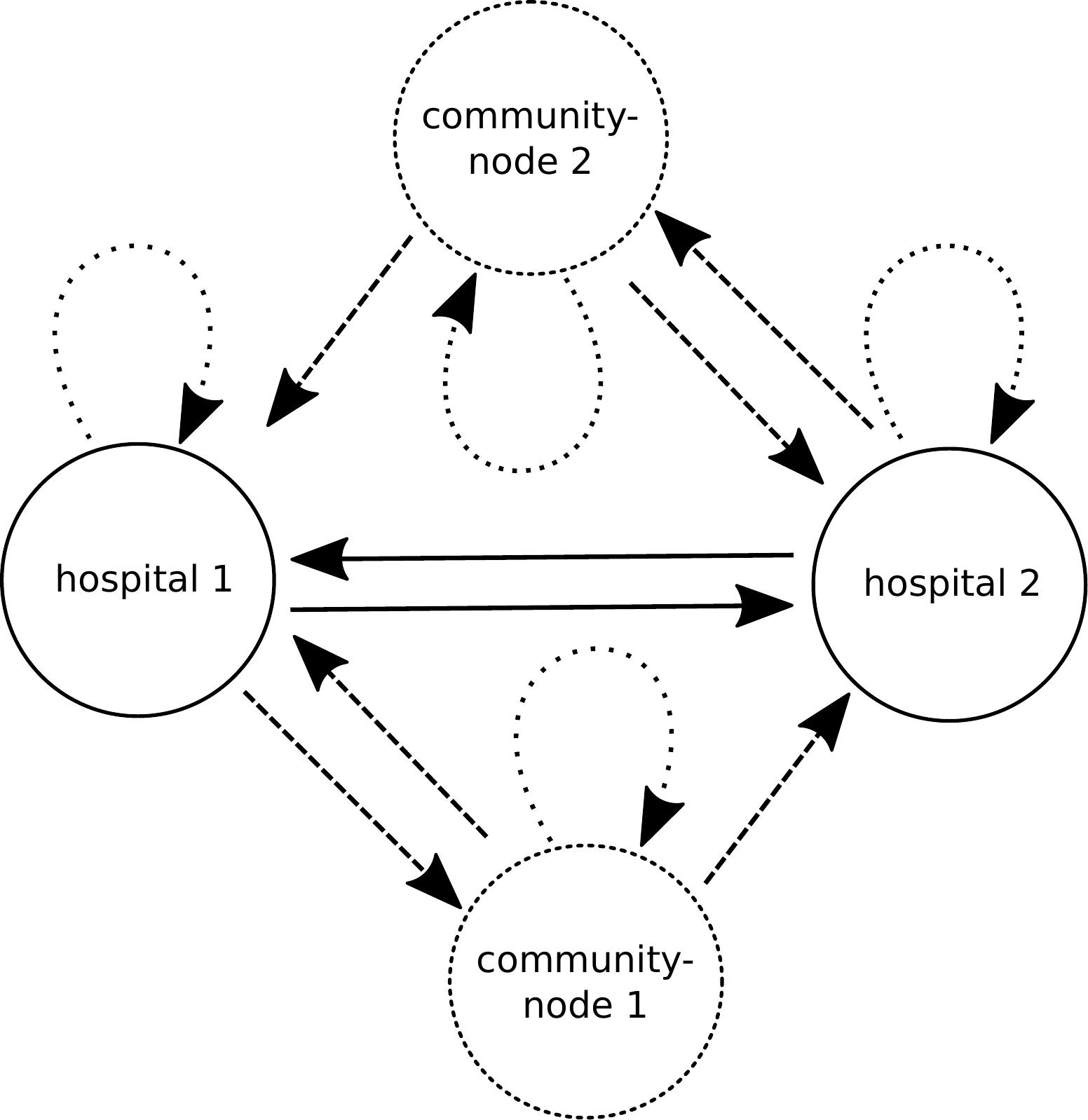}
			\caption{Graphical representation of direct transfers (solid lines) and indirect transfers (dashed lines), in a model comprising two exemplary hospitals. Dotted loops indicate the situations when patients stay in the hospitals/community nodes.
			\label{fig:ghostnodes2}
			}
		\end{figure}

		Below, we describe how we estimate parameters essential for our model from the input provided database. In general, we construct a \emph{directed}, \emph{weighted graph}, based on discharge data, representing the hospital network, where \emph{nodes} correspond to the healthcare facilities and \emph{weights} to transfer probabilities. As patients frequently move not only directly between hospitals, but also they are readmitted after some time period spent outside a hospital, we define additional \emph{community-nodes} (see~Sec.~\ref{sec:soc}), which correspond to the communities associated with the healthcare facilities. Patients leaving a healthcare facility go to such community-node, and then they may be admitted to the same or another facility, c.f.~Figure~\ref{fig:ghostnodes2}. Then, the model allows to simulate in more detail the patient transfers from healthcare facilities to communities and back to healthcare facilities. This way the colonized patients contribute to spread of pathogens in the healthcare system.

		\subsubsection{Extraction of direct and indirect transfers from data}\label{sec:trans:num}

			We distinguish two types of transfers. A \emph{direct transfer} is when patients move from one facility to the next on the same or the next day. In contrast, an \emph{indirect transfer} means a new admission to a hospital after at least one day outside of any hospital. The indirect transfer can be to the same or a different hospital. By a duration of a transfer, we define the length of the period between discharge and next admission. The situation when the patient is re-admitted to the same facility after some time spent in the community is called \emph{(indirect) auto-transfer}.
			
\begin{figure}
				\centering
				\includegraphics[width=0.95\linewidth]{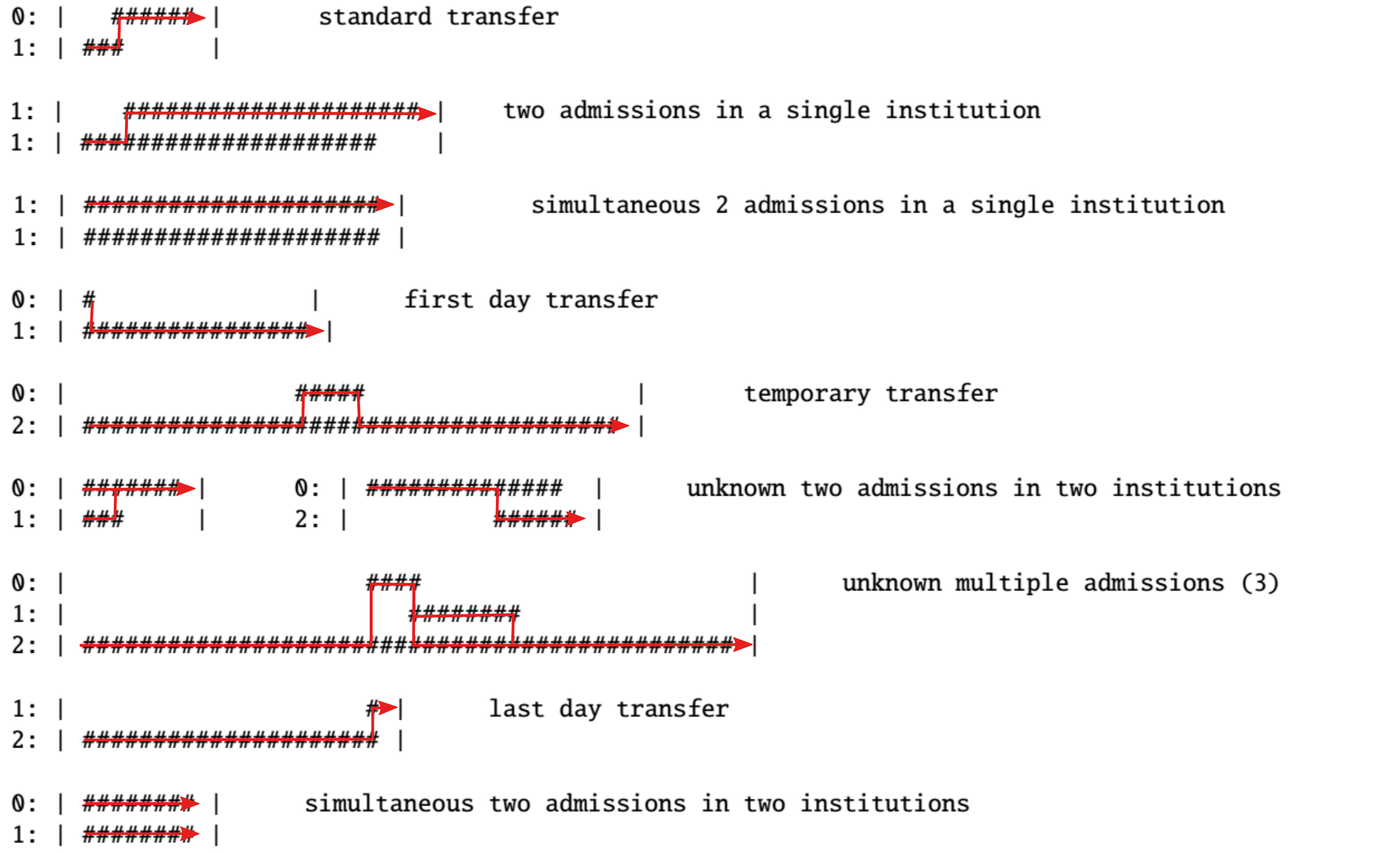}
				\caption{Examples of the results (red line) of the transfer detection algorithm in case of different types of overlaps. Numbers on the left-hand side indicate the numeric codes for particular units, while $\#$ stand for days of stay reported for a given healthcare facilities.
				For the classification of particular types of overlaps see~\cite{Piotrowska2019}.}
				\label{fig:overpal:problem}
			\end{figure}

			The direct transfers can be subject to overlapping (one day or longer) hospital stays, i.e. situation when for the same patient at least two hospitalisations are reported with overlapping periods (c.f. Figure~\ref{fig:overpal:problem}). We detected 304\,833 of these cases for healthcare facilities located in Lower Saxony only. There are several factors contributing to the existence of overlaps. First, the granularity   of admission/discharge is in days, so that a transfer may be indicated by a single-day overlap (112\,368 cases corresponding to 38.1\% of all detected overlaps) as well as two consecutive non-overlapping stays. Moreover, an overlap may be longer than one day if a patient is supposed to return to the originating facility and a bed is kept available for this patient. Overlaps may also occur due to coding errors in the original dataset which cannot be corrected given the anonymized nature of the data. In the context of infection transmission, we are interested in physical and not administrative stays of the patients. For the discussion and analysis of the existing in database overlaps we reefer reader to an open access report~\cite{Piotrowska2019}. 
			
			In order to solve the problem of existing longer than one day overlaps in the dataset, we derived sophisticated algorithm  for the detection of direct and indirect transfers (see the Appendix~\ref{sec:transfer:algorithm} for details).

\begin{figure}
				\centering
				\includegraphics[height=6.5cm]{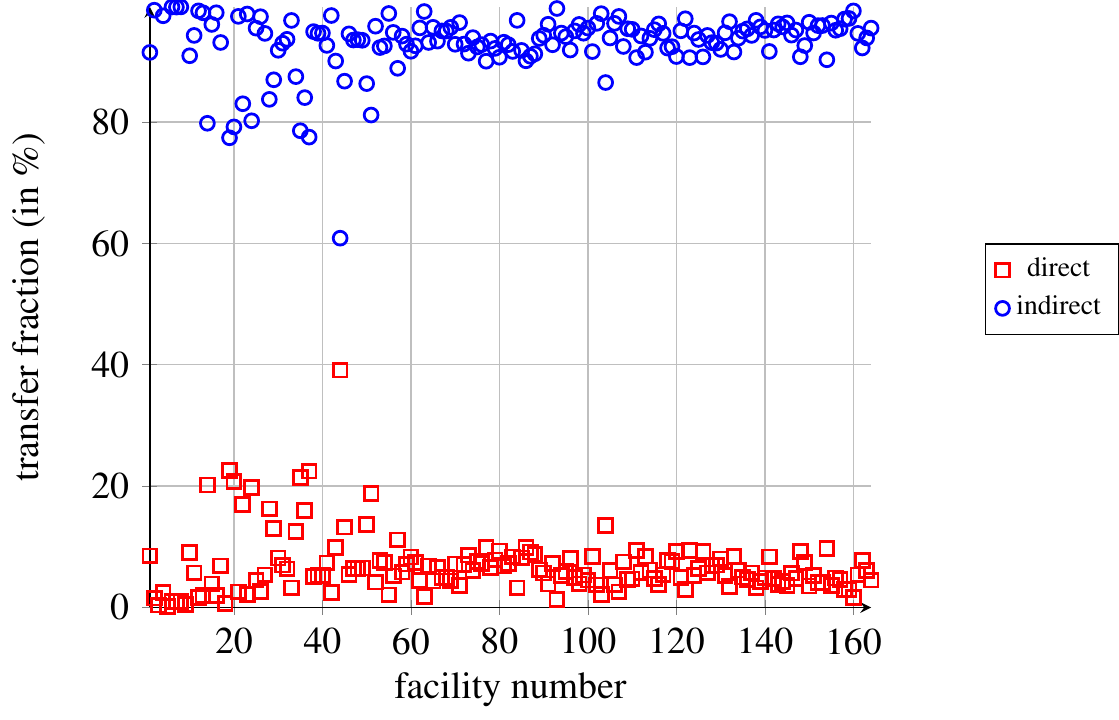}
				\caption{Fractions of direct and indirect (outgoing) (per day) transfers for all healthcare facilities. Healthcare facilities are ordered by the average size of the units, the smallest first. \label{fig:transfers}}
				
			\end{figure}

			Using this  algorithm (see~Section~\ref{sec:transfer:algorithm}), we identified 2\,733\,286 transfers in Lower Saxony, among these 157\,143 were direct transfers (below 6\% of all transfers) and 2\,576\,143 indirect transfers (above 94\%, including 1\,648\,400 auto-transfers).
			In Figure~\ref{fig:transfers} the fractions of both detected types of transfers for each facility are shown. The average length of indirect transfers (stays in the community) was 320.1 days (SD=435.2), while the average length of stays in hospitals was 8.7 days (SD=12.2).
			There are only 573\,045 indirect transfers (about 20\% of all transfers) with a duration of less than 30 days.
			
		\subsubsection{Sizes of hospital nodes}\label{sec:hosp_size}

\begin{figure}[t]
				\begin{subfigure}[t]{0.5\textwidth}
					\centering
					\includegraphics[height=6.2cm]{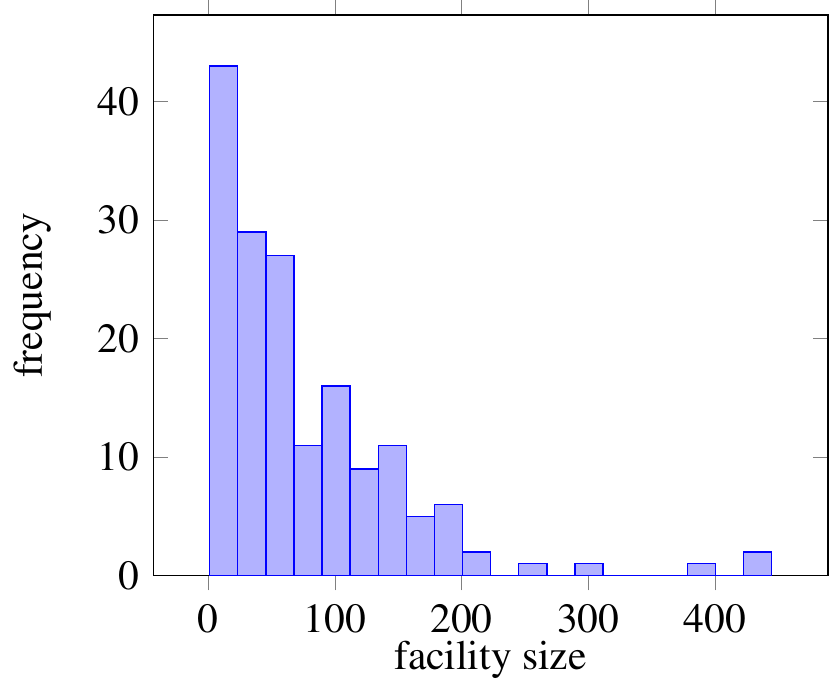}
					\caption{ \label{fig:anal:size:c}}
				\end{subfigure}%
				~
				\begin{subfigure}[t]{0.5\textwidth}
					\centering
					\includegraphics[height=6.2cm]{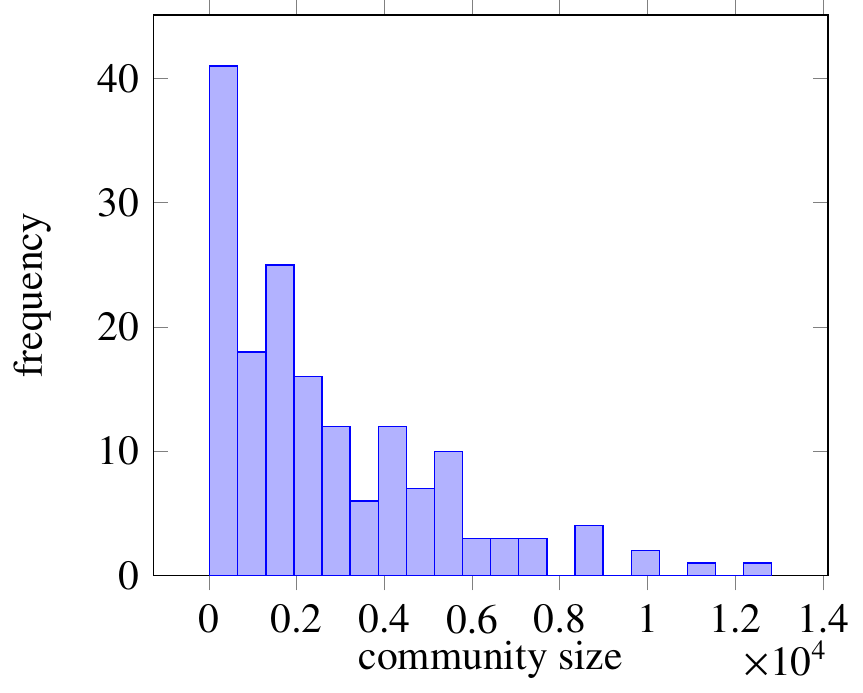}
					\caption{ \label{fig:anal:size:b}}
				\end{subfigure}\\%
				\begin{subfigure}[t]{0.5\textwidth}
					\centering
					\includegraphics[height=6.5cm]{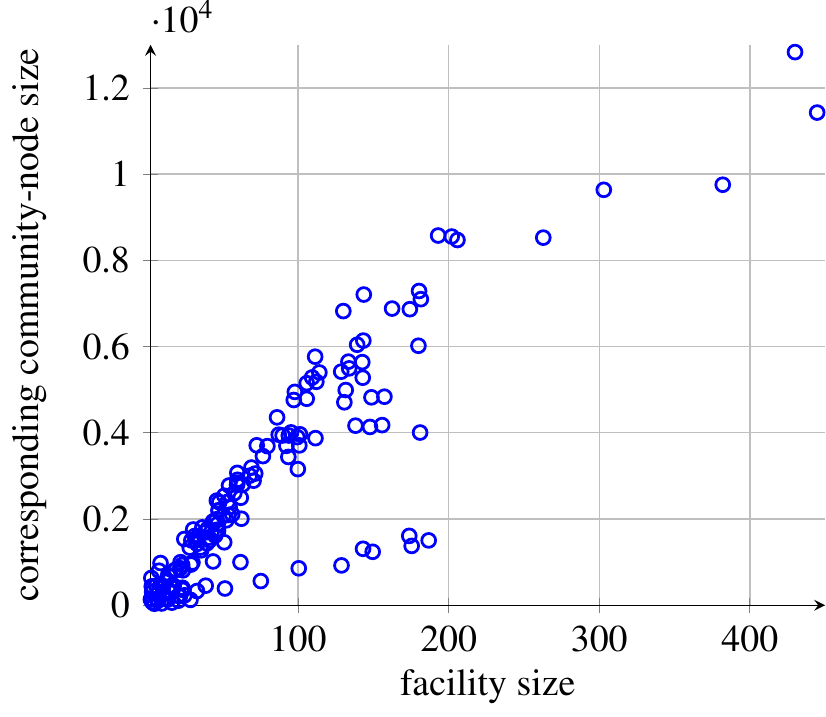}
					\caption{ \label{fig:anal:size:d}}
				\end{subfigure}
				~
				\caption{
				(a) Histogram of estimated hospital sizes, defined as average numbers of the patients in the hospitals within period 2008--2015, calculated for each healthcare facility separately (164 units) --- given the fact that AOK lower Saxony covers 30.52\% of the population in Lower Saxony, the number of beds has to be multiplied by about 3 to get the actual sizes;
				(b) histogram of estimated community sizes, defined as average numbers of the patients in the given community within period 2008-2015, calculated for each  healthcare facility separately, as above the populations have to be multiplied;
				(c) dependence of community-node sizes on corresponding healthcare facilities (164 units).
				\label{fig:anal:size}}
			\end{figure}

			Since healthcare facilities are anonymized in our dataset, we do not have information on their sizes. Moreover, we only have admission and discharge data for a subset of patients, which are the clients of a single insurance company. To estimate the sizes of considered hospital nodes (164 units), we count all patients present on a given day in a given facility, and then take the average over the facility operation period. The distribution of the resulting hospital sizes is presented in Figure~\ref{fig:anal:size:c}.

		\subsubsection{Sizes of community-nodes}\label{sec:soc}
			
			To estimate the average number of people being subject to indirect transfers, we first investigate the change in time of the number of people being in community between two hospitalisations.
			To keep track of the originating hospital of a community dwelling patient, we propose to create a \emph{community-node} for each facility $i$, indexed by $n+i$, where $n$ is the number of all considered healthcare facilities. The patient is assumed to be in the community-node $n+i$ if they were discharged from healthcare facility $i$ earlier and later on (after some time) be admitted to another (possibly the same) healthcare facility $j\in\{1,..., n\}$ or later on did not visit any hospital in the considered period. 

			To estimate the average sizes of community-nodes, we iterate through the dataset period and for every day and for every facility we calculate the number of patients being in indirect transfer, discharged from a given facility waiting for their admission to another one. For $i$ community node we also count patients who are subject to first hospital stay in unit $i$. Then we take  the averages of these numbers.

			In addition, Figure~\ref{fig:anal:size:d} shows the relationship between healthcare facility sizes and sizes of the corresponding community-nodes i.e. populations discharged from healthcare facilities and waiting for (re)admission.

	\subsection{Model description} \label{sec:compmodel}

		We describe the healthcare system as a graph, where the nodes represent healthcare facilities and corresponding community-nodes, with weighted edges representing the probabilities of transfers (both direct and indirect) between the nodes. 

		\subsubsection{Patient traffic}\label{sec:trans} 


			To construct the healthcare network, the model requires input of a list of healthcare facilities ($n$ elements) and corresponding community-nodes ($n$ elements)  with a matrix $A=[A_{i j}]_{i,j=1}^{2n}$ determining the transfer probabilities between the nodes.

			Hospital discharge data were collected for period of $D \in \mathbb{N}$ days, thus by $\mathcal{D} := \{1, \ldots, D\}$ we denote the set of all days for which admission and discharge are registered meaning that every number in $\mathcal{D}$ set corresponds to a single reported day.

			Let $t_{ij}(d)$, $t_{ij}: \mathcal{D} \rightarrow \mathbb{R}^+ \cup \{0\}$ for any $i \neq j$ denote a number of transitions from node $i$ to node $j$, within a single day $d\in\mathcal{D}$, which can be directly extracted from the dataset provided by the insurance company.

			We differentiate four categories of transfers, depending on indices $i,j$:
			\begin{itemize}
				\item $i, j \in \{1, \ldots, n\}$ --- direct transfers between healthcare facilities;
				\item $i\in \{n+1, \ldots, 2 n\}, j \in \{1, \ldots, n\}$ --- indirect transfers from community-nodes to healthcare facilities;
				\item $i\in \{1, \ldots, n\}, j \in \{n+1, \ldots, 2 n\}$ --- transfers from healthcare facilities to community-nodes. They may be also interpreted as initiations of indirect transfers. Since a patient stays in the community-node corresponding to the discharging facility, only $t_{i, n+i} \neq 0$, and for $j \neq n+i$ we have $t_{ij} \equiv 0$;
				\item $i, j \in \{n+1, \ldots, 2 n\}$ --- transfers between community-nodes. Due to the nature of community-nodes, there is no transfer between them, thus these elements are zero except for $t_{ii}$.
			\end{itemize}

			Define $T=[T_{i j}]_{i,j=1}^{2n}$ to be a matrix of the aggregated numbers of all direct and indirect transfers between nodes within a given time period in the following way
			\begin{equation}
			T_{ij} = \sum_{d=1}^D t_{ij} (d).
			\end{equation}
			We refer to each node (hospital or community-node) by their index in the matrix $T$.

			Now, we define $a_{ij}(d)$ to be the per patient probability of a transfer from node $i$ to node $j$ in a single given day $d\in\mathcal{D}$. Then for an arbitrary chosen facility, the average number of transferred patients at a given day is expressed by $a_{ij}(d) p_i$, where by $p_i$ we denote the average length of stay in the $i$-th node, to be defined later. 
			Thus,
			\begin{equation}
				t_{ij}(d) \equiv a_{ij}(d) p_i.
			\end{equation}
			Assuming that the probability of a transfer from node $i$ to node $j$ in a single given day does not depend on choice of a day, i.e. $a_{ij}(d)\equiv A_{ij} = \mathrm{const}$, we get
			\begin{equation}
				\label{eq:tij:obl}
				T_{ij} = \sum_{d=1}^D t_{ij} (d) = D A_{ij} p_i, 
			\end{equation}
			and we determine non-diagonal elements of the matrix $A$ as
			\begin{equation}
				\label{eq:aij:obl}
				A_{ij} = \frac{T_{ij}}{D p_i}=\frac{1}{Dp_i}\sum_{d=1}^D t_{ij} (d).
				\end{equation}
			With elements $A_{ij} := A_{ij}(p_i)$ defined, we may easily determine the diagonal elements $A_{ii} := A_{ii}(p_i)$, corresponding to probability of remaining in the given node as
			\begin{equation}
				\label{eq:A:diag}
				A_{ii} := 1 - \sum_{j \neq i} A_{ij},
			\end{equation}
			and thus matrix $A$ of a Markov chain is determined up to node average populations $\mathbf p = [p_1, \ldots, p_{2n}]$.
			From the mathematical point of view, these values may be set to any positive numbers, provided that $0 \leq A_{ii} \leq 1$.
			In this study, we picked the values of $p_i$ so that the average length of stay of the patients in the nodes agree with the values determined directly from the data.

			Note that the elements on the diagonal of the transfer matrix $A_{ii}$ are the probabilities of stay of patients in $i$ unit, $i=1,..,n$. By the success (at given day $k$) let us denote a situation when a given patient leaves the hospital. It means that for $k-1$ days, with probability of $A_{ii}$ for each day, they were in the hospital and then, at day $k$, they were dismissed with probability $1-A_{ii}$. Thus, we get $A_{ii}^{k-1}(1-A_{ii})$. That is probability mass function of the geometric distribution. Hence, to estimate the average length of patient stays in given hospital $i$ one needs to calculate $1/(1-A_{ii})$.

\begin{figure}
				\centering
				\includegraphics[height=6.5cm]{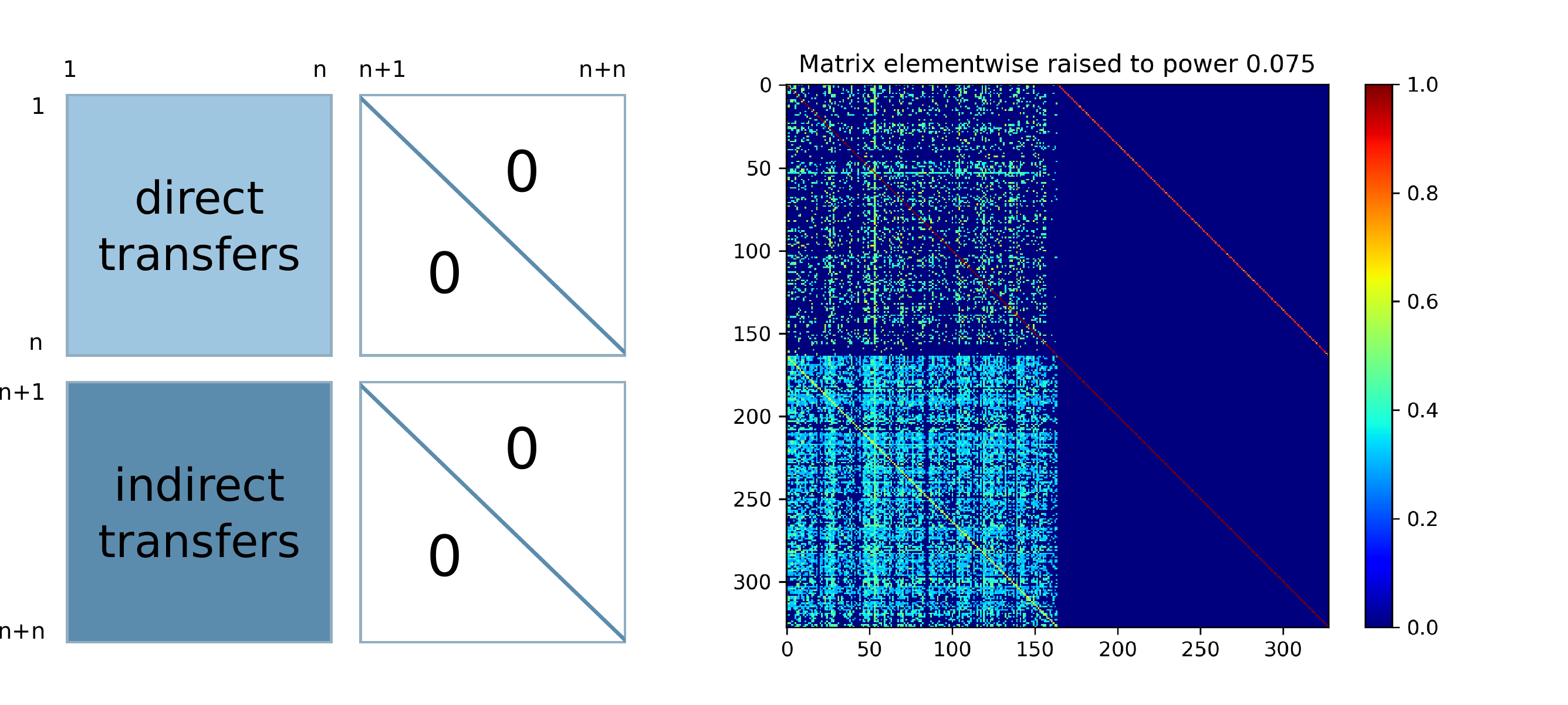}
				\caption{
					Visualization of the transfer probability matrix for $n$ healthcare facilities (numbered from 1 to $n$) and $n$ community-nodes (numbered from $n+1$ to $n+n$), $n=164$: (left) schema presenting four distinguishable blocks; (right) quantitative representation of obtained probabilities, colour pixels denote the probabilities; for the purpose of visualization, all elements were raised to the power 0.075.\label{fig:matrix}
				}
			\end{figure}

			The visualized transfer probability matrix obtained by this procedure can be found in Figure~\ref{fig:matrix}. It has a block structure with the indirect transfer block clearly denser (see discussion in Sec.~\ref{sec:trans:num}). The upper right diagonal block corresponds to discharge from healthcare facilities to community-nodes, and thus it accounts for total indirect transfer quantity. The lower right diagonal block corresponds to patient exchange between community-nodes. Since community-nodes are assumed to be separated, thus only the diagonal elements are present to account for patients remaining at the community-nodes.

			In our model, patient traffic is simulated as follows. Assume that $\mathbf q^0 := [q_1^0, \ldots, q_{2n}^0]$
			is an initial patient probability distribution in healthcare facilities and community-nodes. 
			We calculate the probability changes in the following days by iterating
			\begin{equation}
			\label{eq:continuous:patient:transfer}
			\mathbf q^k :=  \mathbf q^{k-1}A = \mathbf q^0A^k.
			\end{equation}

			If we start with $\mathbf q^0$ being an eigenvector of $A$ (i.e. $\mathbf q^0 A=\lambda \mathbf q^0$, where $\lambda$ is an eigenvalue of matrix $A$), we get $\mathbf q^k = \lambda^k \mathbf q^0$.
			Moreover, if matrix $A$ represents a {\it regular} Markov chain (i.e. there exists $k$ such that all elements of $A^k$ are positive), then we can find the stationary probability distribution $\mathbf w$ of the patients in hospitals for our network model by solving system $ \mathbf w A= \mathbf w$, since in such a case $\lambda=1$ is a eigenvalue of matrix $A$. On the other hand, if the considered Markov chain is {\it absorbing} we then would be able to calculate what is the probability of absorption from one state to another and the average number of steps before the absorption happens.
			It is natural to expect that Markov chain describing patient transfer process is not absorbing, as obviously no such phenomenon is observed in real healthcare systems.

\begin{figure}[t]
				\begin{subfigure}[t]{0.5\textwidth}
					\centering
					\includegraphics[height=6.5cm]{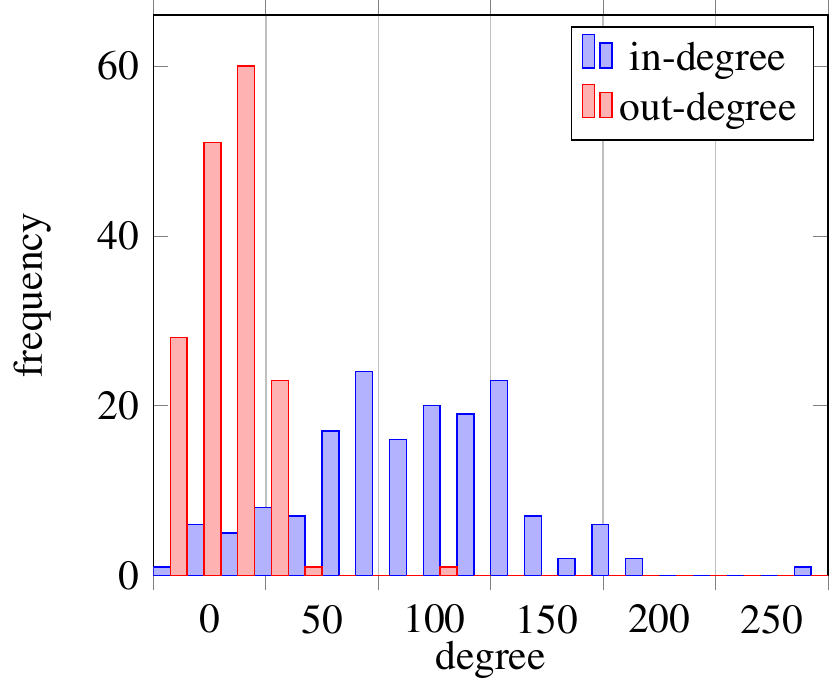}
					\caption{ \label{fig:net:char:a}}
				\end{subfigure}%
				~
				\begin{subfigure}[t]{0.5\textwidth}
					\centering
					\includegraphics[height=6.5cm]{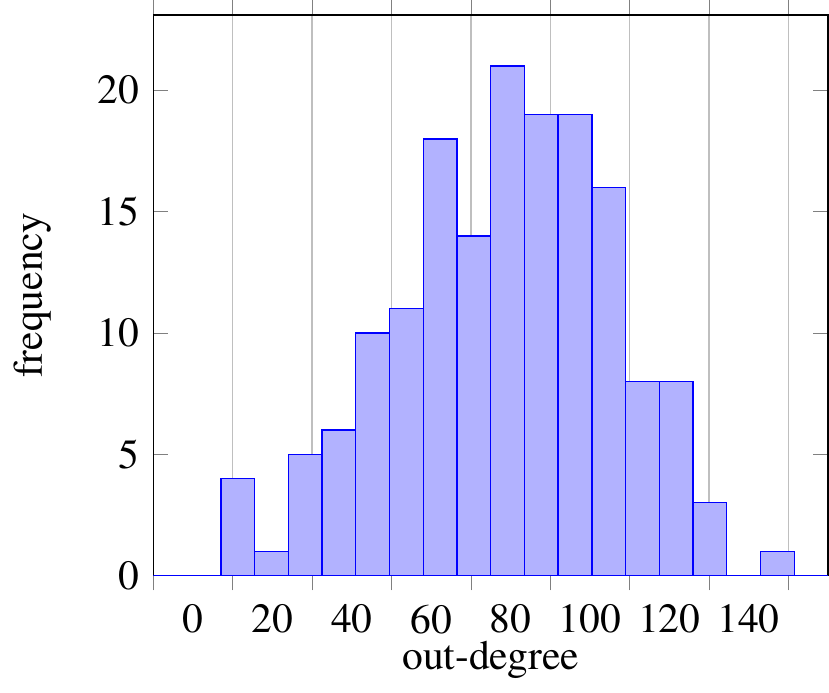}
					\caption{ \label{fig:net:char:c}}
				\end{subfigure}
				\caption{
					(a) In-degree and out-degree for each hospital node of the network (self-loops not included);
					(b) out-degree for each community node of the network (self-loops not included), in-degree for all community nodes are equal to 1.
					\label{fig:net:char}}
			\end{figure}

			In our analysis, we discuss mostly the stationary distribution of patients counted as a number of patients in a given healthcare facility, while in the Markov processes we consider the stationary probability distribution so that the vector of probabilities of patients is in given units. However, we can easily interchange these vectors by dividing/multiplying them by the number of patients in the healthcare system. In case of the network derived from data of Section \ref{sec:description:of:data:set}, the regularity of matrix $A$ was checked numerically by empirical verification ($A^5$ has positive elements only), but this result is dependent on numerical errors. Nevertheless, we can use a lemma providing an analytical argument, for details see \cite{Lonc2019} or Appendix~\ref{sec:app}. A non-zero pattern of matrix $A$ is presented in Figure~\ref{fig:matrix}. Then, in Figure~\ref{fig:net:char}, we present in- and out-degree of all network nodes, calculated using transfer probability matrix $A$.

		\subsubsection{Pathogen transfer dynamics}\label{sec:continuous:susceptible:infected} 

			To model the spread of pathogens within nodes, we use a \emph{susceptible-infectious-susceptible model} (SIS),~\cite{Martcheva2015}. In this classic approach, we assume that we have two well mixed and constant in time populations of $P$  individuals that can be susceptible or infectious. Let $S_f(t)$ and $I_f(t)$ denote the fraction of susceptible and infectious individuals at time $t$, respectively, then by term $\beta S_f(t)I_f(t)$ we describe the infection process due to the contact of susceptible and infectious individuals, while by $\gamma I_f(t)$ we include a recovery process, so that we obtain a system of the following ordinary differential equations
			\begin{equation}\label{eq:SIS:frac}
			\begin{split}
			\frac{d }{d t}S_f(t) & = - \beta  S_f(t) I_f(t) + \gamma  I_f(t),
			\\ 
			\frac{d }{d t}I_f(t) & = \beta  S_f(t) I_f(t) - \gamma  I_f(t),
			\end{split}
			\end{equation}
			where $S_f+I_f=1$ due to the definition of $S_f$ and $I_f$,
			and as a consequence
			\begin{equation}\label{eq:frac:I}
			\frac{d }{d t}I_f(t)  = -\beta (I_f(t))^2 + (\beta -\gamma )I_f(t),
			\end{equation}
			which is a logistic equation, \cite{Verhulst1838}, with analytical solutions.
			Solutions to Eq.~\eqref{eq:frac:I} exist globally and are unique. Moreover, Eq.~\eqref{eq:frac:I} has two steady states: 
			a trivial one (locally stable for $\beta<\gamma$ and unstable for $\beta>\gamma$) and positive one providing  $\beta>\gamma$ (which is stable whenever it exists). 	
			Thus, for $\beta\leq \gamma$ the fraction of infectious individuals goes to zero, while for $\beta>\gamma$ fraction of infectious individuals stabilizes at a certain level.

			In our case, in nodes, it would be favourable to operate on the number of susceptible and infectious patients $S, I$ instead of the fractions $S_f, I_f$. Hence, later on we consider
			\begin{equation}\label{def:SI}
			S(t)=S_f(t)\cdot P, \quad I(t)=I_f(t)\cdot P,
			\end{equation}
			where $I_f(t)$ and $S_f(t)$ are solutions to~\eqref{eq:SIS:frac} and $P$ denotes the population size in the  node. For the MRSA example used in this study, the "infectious" compartment corresponds to individuals currently colonized as well as to those with manifest MRSA infection and we do not distinguish those two populations further.

			We introduce this simple SIS model into all healthcare facilities in our network. In community-nodes, we use the same model but assume no transmission ($\beta = 0$) as in this work, we focus only on hospital acquired infections. As a consequence in the following, we assume that $\mathbf p^k = \mathbf s^k + \mathbf i^k$, $\mathbf s^k := [s_1^k, \ldots, s_{2n}^k]$, $\mathbf i^k := [i_1^k, \ldots, i_{2n}^k]$, where as before $\mathbf p^k=[p_1^k,...,p_{2n}^k]$ denotes the average distribution of the patients in network nodes after $k$ iterations, while $\mathbf i^k$ and $\mathbf s^k$ stand for average distributions of infectious and susceptible patients in nodes, respectively.

			Now let us define the numerical subroutine \texttt{sisolve} as follows. For a given $S_0$, $I_0$, $t_0$, $t_1$, $\beta$, $\gamma$, it returns values $S_1$, $I_1$ (i.e. the numbers of susceptible and infectious individuals after time $t_1-t_0 = \SI{1}{day}$, respectively), where
			\begin{equation}
			S_1 := S(t_1),
			\quad
			I_1 := I(t_1),
			\end{equation}
			where functions $S,I$ are the solutions 
			of the SIS model (with initial condition $S(t_0) = S_0$ and  $I(t_0)= I_0$) given by~\eqref{def:SI} .
			So basically $\mathtt{sisolve}: \mathbb{R}^6 \rightarrow \mathbb{R}^2$ together with
			the SIS model defines our modelling approach as follows:
			\vspace{0.5cm}
			\begin{algorithmic}[1]
				\Require $\mathbf s^0 = [s_1^0, \ldots, s_{2n}^0]$, $\mathbf i^0 := [i_1^0, \ldots, i_{2n}^0]$, $T \in \mathbb{N}$;
				\State $t:=0$;
				\State $\mathbf s = \mathbf s^0$;
				\State $\mathbf i = \mathbf i^0$;
				\While{$t < T$}
				\ForAll{$j \in \{1, \ldots, 2n\}$} \label{algo:csi:innerfor}
				\State $(s_j, i_j) := \mathtt{sisolve}(s_j, i_j, t, t+1, \beta, \gamma)$;
				\EndFor
				\State $\mathbf s := \mathbf s A $; \label{algo:csi:As}
				\State $\mathbf i := \mathbf i A $; \label{algo:csi:Ai}
				\State $\mathbf s^t := \mathbf s; \mathbf i^t:= \mathbf i$; \label{algo:csi:results}
				\Comment{These are the results}
				\EndWhile
			\end{algorithmic}
			\vspace{0.5cm}
			Loop~\ref{algo:csi:innerfor} can be parallelized. This is also true for multiplications~\ref{algo:csi:As}, \ref{algo:csi:Ai}, but here some inter-process information exchange is necessary and thus for our designed numerical procedure there is no need of having the same recovery and transmission rates for all healthcare facilities and/or community-nodes.
			The results of the described algorithm, i.e. distribution of susceptible and infectious patients versus discrete time steps, are computed in line~\ref{algo:csi:results}.

			We assume that the time-step is one day, and that the recovery and transmission rates do not vary between the facilities. This behaviour may be easily modified if necessary.
			We also assume that the transfer probabilities for infectious patients and susceptible patients are the same (lines \ref{algo:csi:As}, \ref{algo:csi:Ai}). To change this assumption, we will have to generate different Markov Chain matrices for these groups, but otherwise the presented algorithm remains unchanged.


		%
		%

	\subsection{Estimation of parameters for transmission and recovery}\label{sec:modelparam}

		For the purpose of this paper, we use  as reference parameters values already reported in the literature for \emph{methicillin-resistant Staphylococcus aureus} (MRSA). If necessary, we vary the reference values to  analyse in-depth the behaviour of our model.
		The recovery rate $\gamma$ is equal to one over mean time spent by an individual in the infectious class and it is measured in units of day$^{-1}$. Following Scanvic~et~al.~\cite{Scanvic2001} and Donker~et~al.~\cite{Donker2010, Donker2014},  we assume that the mean time of  MRSA colonization is 365 days leading to   $\gamma=1/365$~day$^{-1}$. In addition, we assume that recovery rates in particular facilities do not depend on the patient characteristics or healthcare facility characteristics itself. Thus, for all nodes representing healthcare facilities, we use the same value for the recovery rate. The same holds for the all community-nodes if not indicated otherwise. To determine the reference transmission rate value is not an easy task. In papers focusing particularly on MRSA transmissions and using SIS model to mimic bacteria  transmission like \cite{Donker2012,Donker2010} this parameter was not reported. On the other hand, in \cite{Donker2014} parameter $\beta$ was set at 0.085, 
		while in~\cite{Donker2017} wild range of the transmission parameters were tested, cf.~Table 1 in~\cite{Donker2017}. 
		

		We additionally assume that in the community-nodes representing community patients can not get colonized and thus transmission rates for the community-nodes are set to zero~(Table~\ref{tab:sis:params}). For the set of simulations presenting the impact of the SIS model parameters on the dynamics of the whole model we vary both SIS parameters: recovery rate and transmission rate, as reported in~Table~\ref{tab:sis:params} set type MRSA.

		For the set of simulations investigating the impact of indirect transfers, we start with a specific parameter combination, $\gamma_h := \gamma_c = 1/365 \;\mathrm{day}^{-1}$ and $\beta=0.06 \;\mathrm{day}^{-1}$, 
		resulting in stabilization of the system-wide community prevalence within period of 7000 days at the level of 6.7\%, and stabilization of the system-wide hospital prevalence at the level of 17.8\%. The latter prevalence is close to the mean prevalence reported in \cite{Donker2010, Donker2012, Donker2014} for considered bacteria.

		In subsequent simulations, we model the effect of increased indirect-transfer pathogen transmission between the facilities by investigating the wide spectrum of $\gamma_c$ parameters. We start from the value $0.125/365 \;\mathrm{day}^{-1} \approx 3.42 \times 10^{-4} \;\mathrm{day}^{-1}$ corresponding to much lower (than the one proposed in \cite{Scanvic2001,Donker2010, Donker2014}) recovery in community and then  we gradually increase the recovery rate in the community up to $4096 / 365 \approx \;\mathrm{day}^{-1} 11\;\mathrm{day}^{-1}$, which gives mean duration of recovery of about two hours. Using this approach, the movement of patients in the system remains the same, but the patients recover faster in the community, so that pathogen transmission through this channel is reduced, and in the most extreme case ($\gamma_c = 4096 / 365  \;\mathrm{day}^{-1} \approx 11 \;\mathrm{day}^{-1}$) it is virtually disabled. In this case, the transmission is limited to the hospital environment only.

		\begin{table}
			\centering
			\small
			\caption{\label{tab:sis:params}SIS parameters used in simulations.}
			\begin{tabular}{l|l|l|l|l}
			 Set type&	Parameter & Reference values & Unit  &  Reference \\
				\hline
			\multirow{3}{2cm}{MRSA} &	Recovery rate ($\gamma$)	& 1/365 &day$^{-1}$  & \cite{Scanvic2001,Donker2010, Donker2014}\\ 
			\cline{2-5}
				& Transmission rate in hospital ($\beta$) &     0.06 &day$^{-1}$  &  see Section~\ref{sec:modelparam}\\ 
			\cline{2-5}
				& Transmission rate in community ($\beta$) &     0   &day$^{-1}$  & assumed\\
				\hline
			\multirow{3}{2cm}{Parameter analysis} &	Recovery rate ($\gamma$)	& 0.5/365--8/365 &day$^{-1}$  & assumed\\ 
			\cline{2-5}
				& Transmission rate in hospital ($\beta$) &     0.04--0.85  &day$^{-1}$  & assumed\\ 
			\cline{2-5}
				& Transmission rate in community ($\beta$) &     0   &day$^{-1}$  & assumed\\
				\hline
			\multirow{3}{2cm}{Indirect transfer impact}&	Recovery rate in hospital ($\gamma_h$)	&  1/365 & day$^{-1}$  &  \cite{Scanvic2001,Donker2010, Donker2014} \\
			\cline{2-5}
			& 
			Recovery rate in community ($\gamma_c$)	&  0.125/365--4096/365 & day$^{-1}$  &  see Section~\ref{sec:modelparam}\\
			\cline{2-5}
			& 
				Transmission rate in hospital ($\beta$) &     0.06   & day$^{-1}$  & see Section~\ref{sec:modelparam}\\
			\cline{2-5}
			& 
				Transmission rate in community ($\beta$) &     0   &day$^{-1}$  &  assumed\\
			\end{tabular}
		\end{table}

	\subsection{Numerical simulation setup}
		For simulations of the pathogen spread in the healthcare network, we use software developed by K.~Sakowski and M.J.~Piotrowska (for details and documentations see~\cite{emergenetpackage}).
		The software is written in Python and comprises two main modules: one for patient transfers between healthcare facilities and one for pathogen spread within healthcare facilities. The former module is superior
		to the latter one in the sense that it governs the simulation flow. Moreover, it performs the parallelization through the MPI library.

		In the first step, the healthcare facility nodes (and their community-nodes, if applicable) are distributed between the available processors. Next, the transfer matrix, described in detail in Section~\ref{sec:trans}, is divided into submatrices, corresponding to blocks of nodes given to the subsequent processes; then these blocks are also distributed to the corresponding processors.
		The internal model, responsible for calling independent instances of the intra-facility model for pathogen spread simulation with a SIS model as defined in Section \ref{sec:continuous:susceptible:infected}, is then initialized.
		After the intra-facility simulations (covering one day period) are finished, the patients are being transferred between nodes and this process is repeated for the next day, and so on.

	\subsection{Simulation plan}

		Using static network structure described above and the SIS transmission model, we first assess how MRSA spreads in the network (including direct and indirect transfers) and how it depends on the parameters describing dynamics of the spread of the bacteriae within healthcare facilities and corresponding community. We also check if the starting point of spread of infection strongly influences model dynamics.

		A particular focus is placed on the role of transmission parameters and the initial infection point for the behaviour of the system until a steady state is reached, and on the prevalence of MRSA colonization during steady state in individual healthcare facilities and their corresponding community nodes. By increasing the recovery rate in community-nodes from  \SI{0.125/365}{day^{-1}} 
		to about \SI{4096/365}{day^{-1}} (instant recovery in the community), we investigate the effects of stepwise restricting the contribution of indirect transfers to pathogen spread.

\section{Results}

	\subsection{Impact of the initial infection point}\label{sec:inital:infection}


		Let us consider SIS reference parameters for MRSA in Table~\ref{tab:sis:params}.
		Independently of the initial starting point of pathogen spread, we observed three phases of the spread of MRSA in our network.  In the initial phase, the overall prevalence was small, close to zero. With time prevalence increases slowly, mainly in the initially colonized facility. Then, there is a transition phase, where the number of colonized patients increases rapidly. The increase in prevalence in this phase is similar to the behaviour of an arctan function near zero. Then, the final prevalence is reached in individual facilities and overall. The proportion of infectious individuals does not change any more, so that a final steady state is reached.

\begin{figure}
			\centering
			\includegraphics[width=1\linewidth]{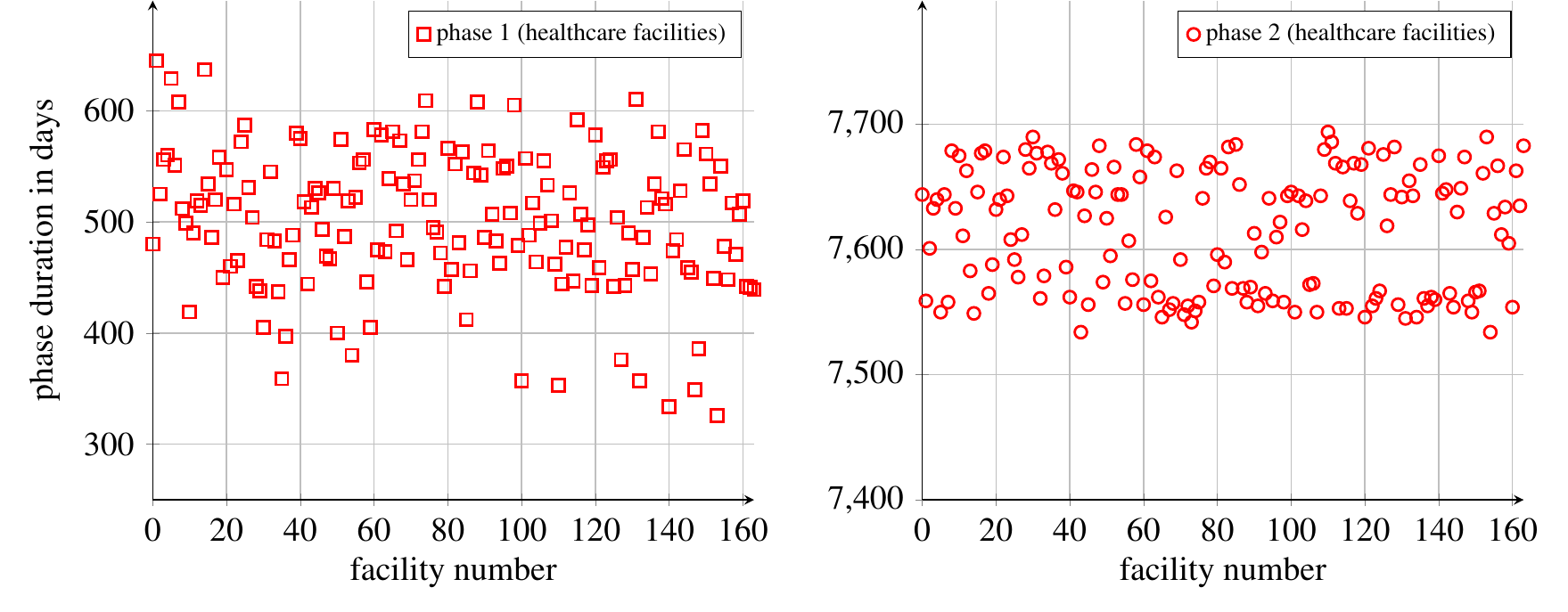}
			\caption{
				Average system-wide phase durations expressed in days versus healthcare facility which was selected as the initial infection point (sorted by avg. hospital size, with the smallest first).
				Phase 1 denotes beginning of the epidemic (prevalence lower than 10\% of the final prevalence); phase 2 denotes transition state (prevalence < 99.9\% final prevalence).
				The phase durations correspond to a system-wide prevalence, not to individual facility prevalences.
				Infection was started by a single infectious patient originating from a given facilities.
				\label{fig:ghinitphase-duration}
			}
		\end{figure}

		These final states are similar, independently of the initially infected healthcare facility (see Figure~\ref{fig:ghinitphase-duration}). These results suggest that there is a single final state, corresponding to some stationary state of the simulated system.
		The time to reach the final state depends on the initial conditions, but differences are very small. The initial phase lasts about 400-600 days, and then the transition phase takes up to 7700 days (see Figure~\ref{fig:ghinitphase-duration}).

\begin{figure}
			\centering
			\includegraphics[height=6.5cm]{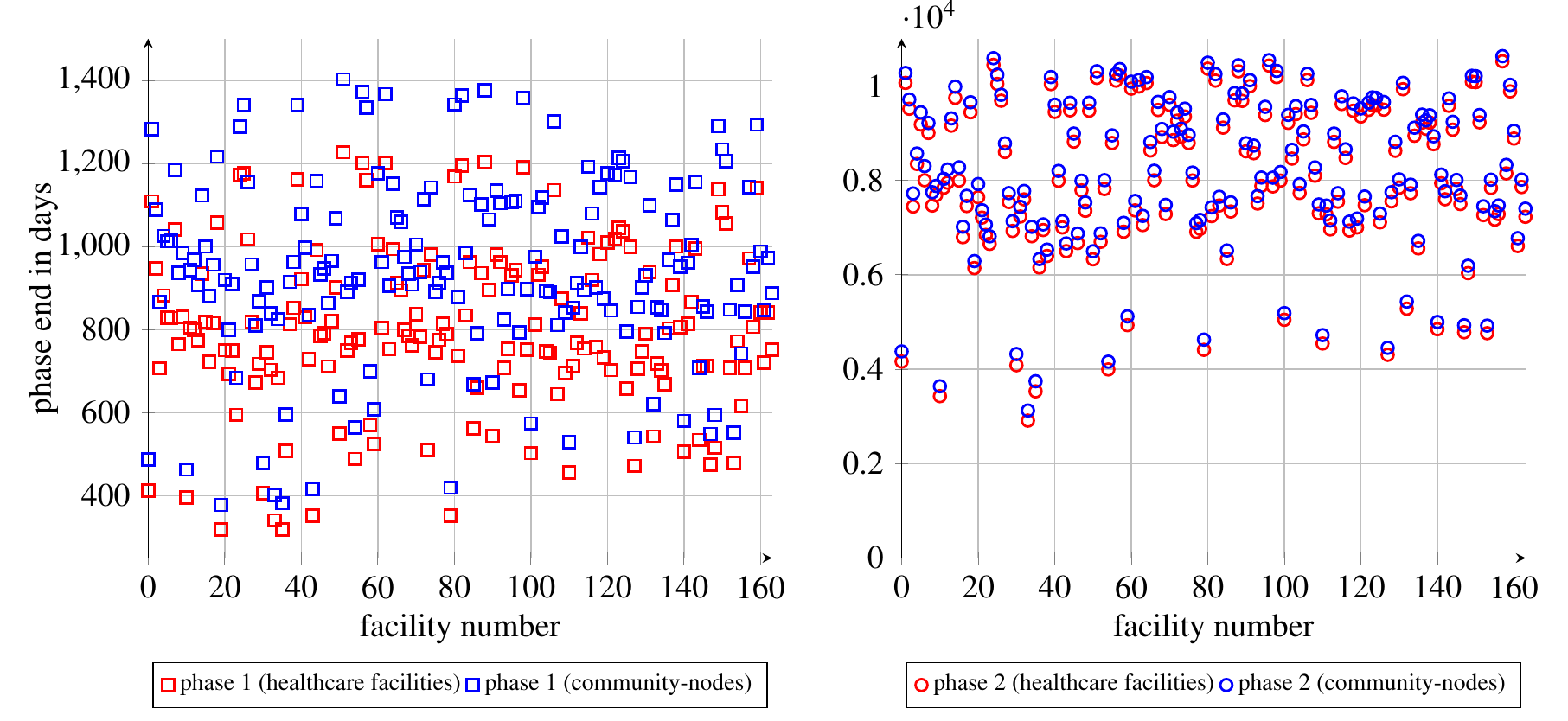}
			\caption{Individual phase ends in healthcare facilities and in corresponding community-nodes. Presented results are averaged-out amid all possible initial infection points.}
			\label{fig:ghnghphase-end}
		\end{figure}

\begin{figure}
			\centering
			\begin{subfigure}[t]{0.49\textwidth}
				\centering
				facilities\\
				\includegraphics[width=\textwidth]{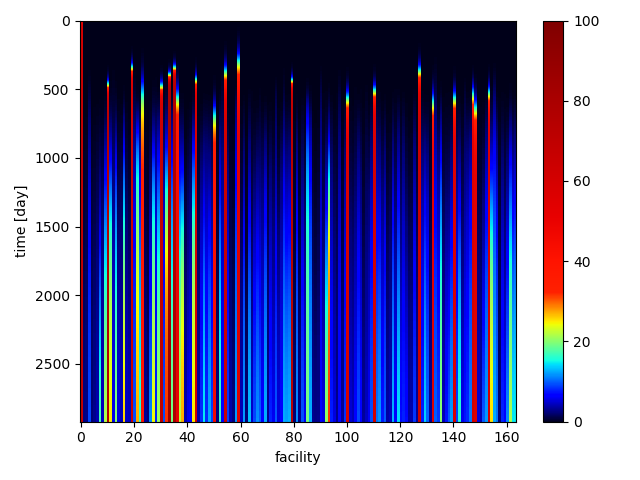}
			\end{subfigure}
			\begin{subfigure}[t]{0.49\textwidth}
				\centering
				community-nodes\\
				\includegraphics[width=\textwidth]{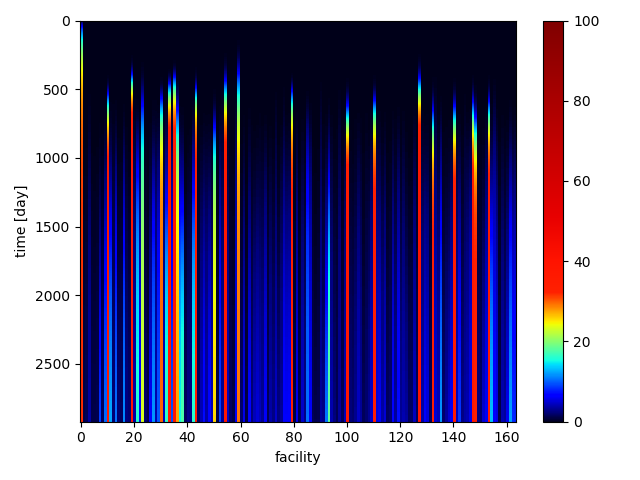}
			\end{subfigure}
			\begin{subfigure}[t]{0.49\textwidth}
				\centering
				\includegraphics[width=\textwidth]{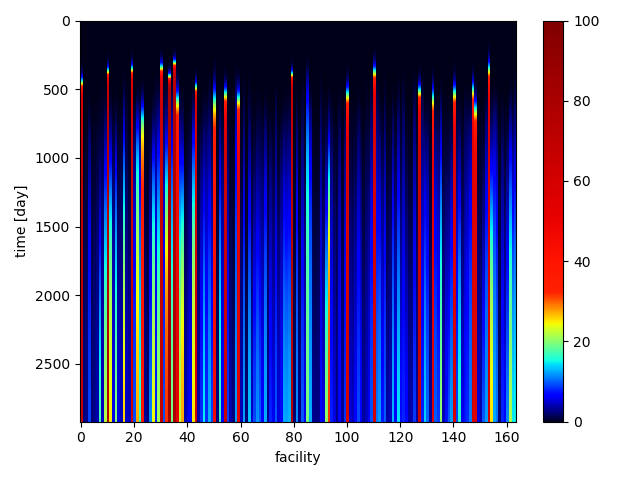}
			\end{subfigure}
			\begin{subfigure}[t]{0.49\textwidth}
				\centering
				\includegraphics[width=\textwidth]{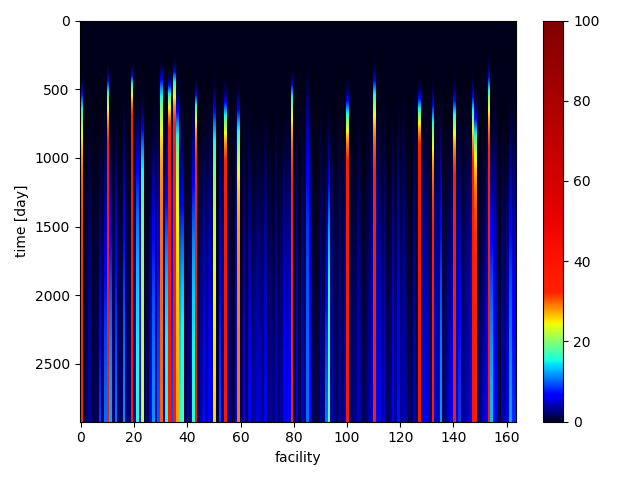}
			\end{subfigure}
			\caption{Network prevalence (over time) in percentage of infectious individuals per healthcare facility (left column) and corresponding community-node (right column).
				Deep blue corresponds to low infection proportions (lower than 20\%); yellow or red to facilities with high infection proportions (> 50\%).
				Healthcare facilities are ordered by average size, with the smallest first.
				The process was started by a single infectious patient located in facility number: 
				(upper row)	1  (the smallest); (lower row) 164  (the biggest). 
				\label{fig:infection:rates}
			}
		\end{figure}

\begin{figure}
			\centering
			\includegraphics[height=6.5cm]{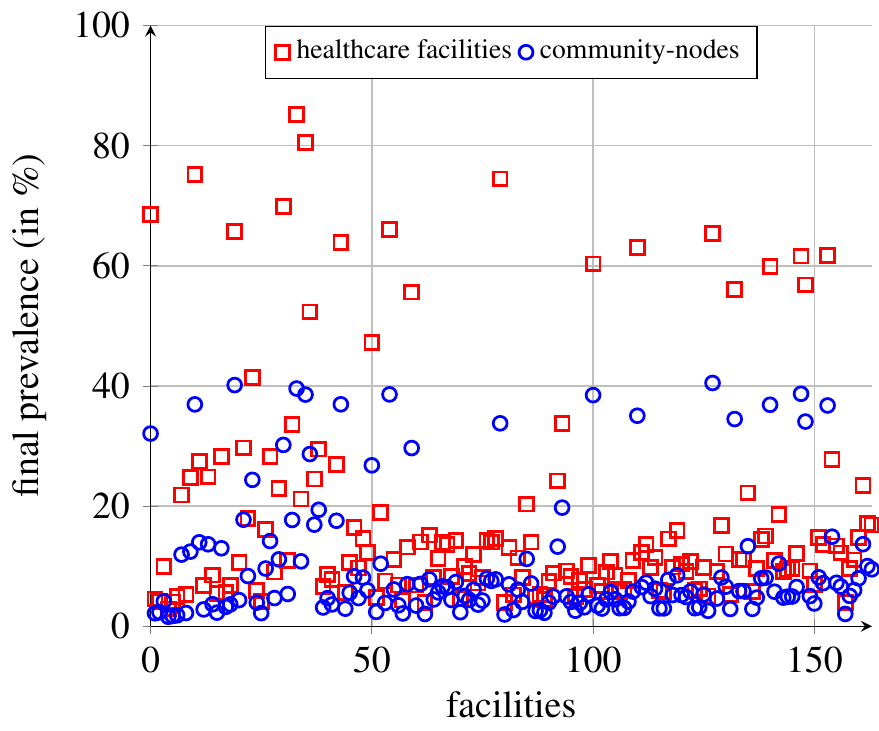}
			\caption{Comparison of the final percentage of infectious individuals in given hospitals/community-nodes. Healthcare facilities are ordered by the average size of hospitals, the smallest first. Transmission was started by a single infectious patient originating from a given facility, and the results are averaged out through all initial facilities. However, differences of the final prevalence between initial originating facilities are negligible.
				\label{fig:prevalencegnngnsort}
				}
		\end{figure}

		If we focus on the individual facilities (see Figures~\ref{fig:ghnghphase-end} and~\ref{fig:infection:rates}),  there is a high heterogeneity in lengths of the studied phases. Some facilities reach their final state within 4000 days; the mean duration of initial phase is, however, at more than 10000 days. These patterns do not depend on the initial conditions under consideration (except for the initial infection points, where the spread is faster). Moreover, there is no clear association between size of a facility and final prevalence or time to reach it. However, lower final prevalence generally corresponds to longer transition phase.
		The final prevalence in the community is generally much lower than in the healthcare facilities. There is a strong association between the combined length of initial/transition phases for healthcare facilities and their corresponding community-nodes. Analogous patterns can be observed for individual facility/community-node combinations --- the final prevalence for individual community-nodes is generally much lower than the final prevalence of the corresponding facility (Figure~\ref{fig:prevalencegnngnsort}).

		\FloatBarrier

	\subsection{Influence of SIS parameters on the final prevalence}
		\label{sec:sis:influence}
\begin{figure}
			\centering
			\begin{subfigure}[t]{0.49\textwidth}
				\centering
				\includegraphics[width=1\textwidth]{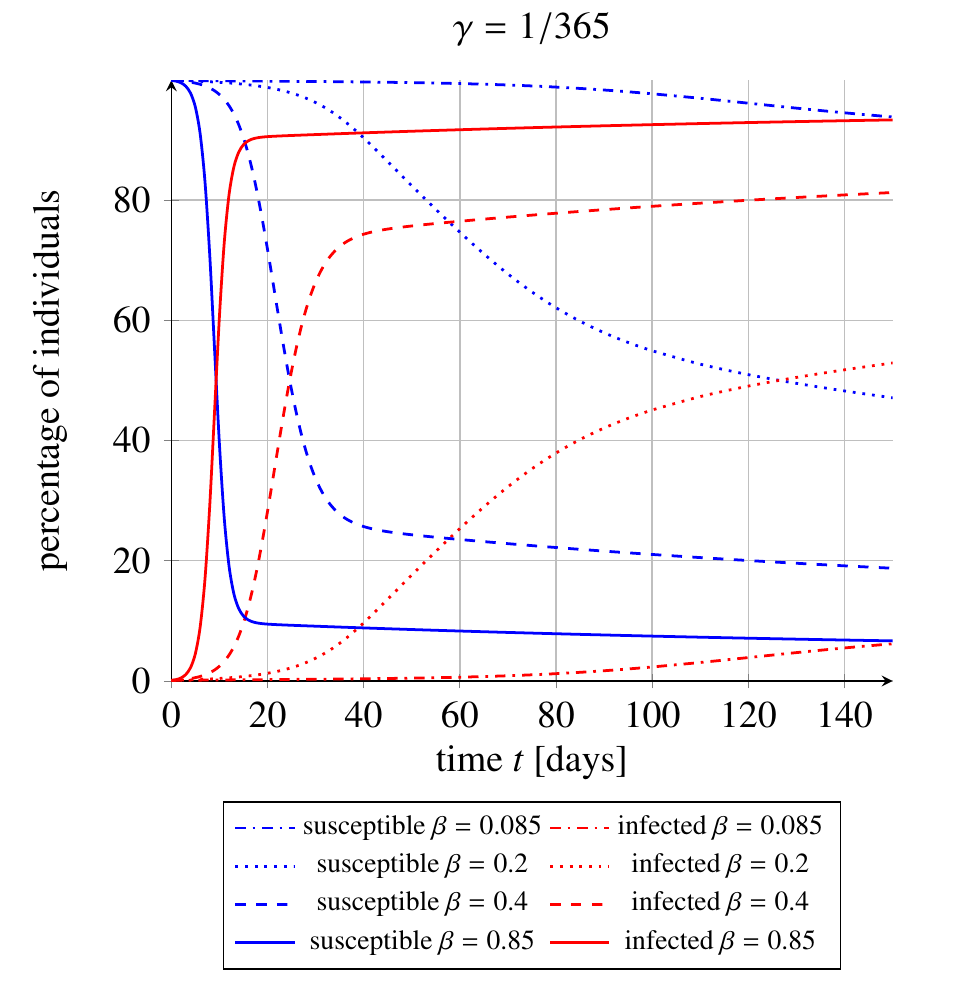}
				\caption{ \label{fig:prevelence:beta:a}}
			\end{subfigure}
			\begin{subfigure}[t]{0.49\textwidth}
				\centering
				\includegraphics[width=1\textwidth]{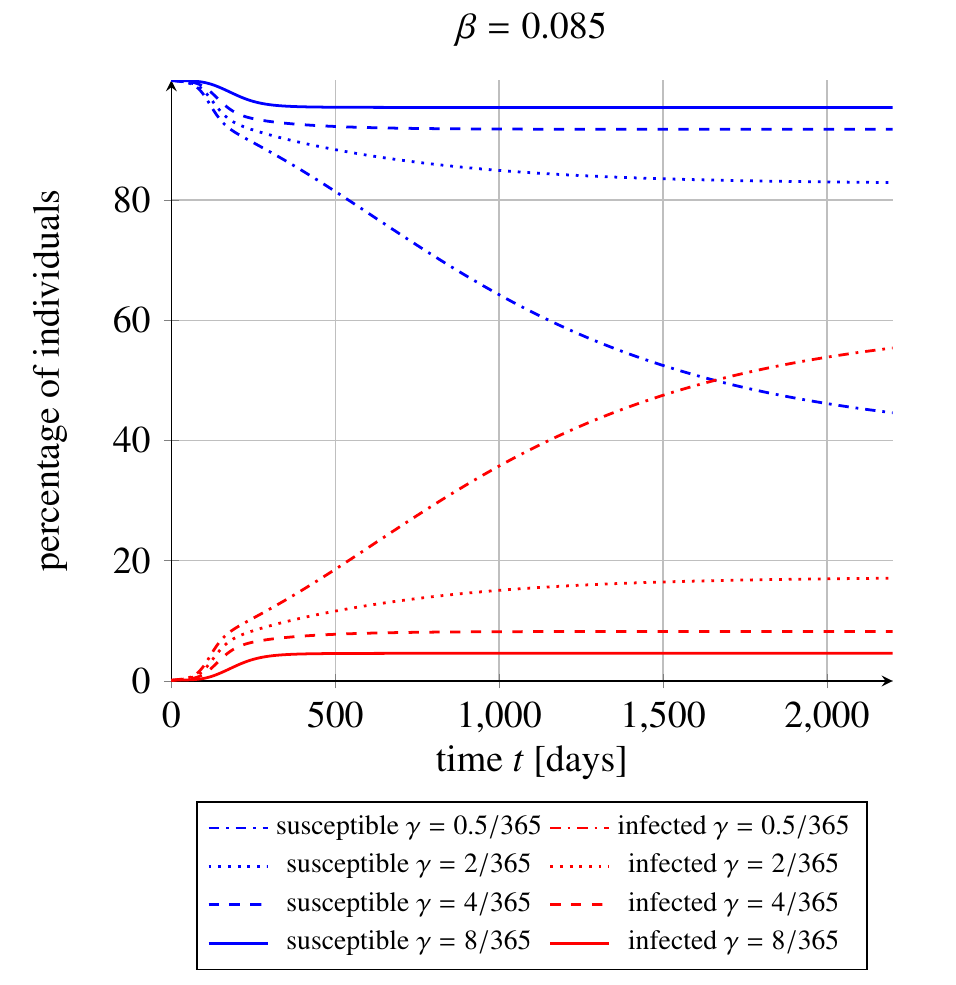}
				\caption{ \label{fig:prevelence:gamma:b}}
			\end{subfigure}
			\caption{
			(a) Effect of the change of the transmission rate on the percentage of susceptible (blue) and infective (red) individuals in the healthcare facilities network. (b) Effect of the change of the recovery rate on the percentage of susceptible (blue) and infectious (red) individuals in the healthcare facility network. Spread of infection was started by a single infectious patient located in a 
				(left) small, 
				(right) big 
				facility.
				}
			\label{fig:prevelence}
		\end{figure}

		In Figures~\ref{fig:prevelence:beta:a}--\ref{fig:prevelence:gamma:b}, we present the effect of variation in transmission rates and recovery rates in the SIS model on the cumulative proportion of all susceptible (blue) and infectious (red) individuals in healthcare facilities (counted at one point) for the sets of parameters being  perturbations of the SIS reference values reported in Table~\ref{tab:sis:params} (set type: MRSA) as indicated in legends. For all presented simulations, infection was initiated by the introduction of uniformly distributed 0.1\% infectious patient in whole population.

		In a model accounting for both direct and indirect transfers, our results (Figure~\ref{fig:prevelence}) indicate that both parameters, the transmission rate $\beta$ and the recovery rate $\gamma$, impact the final network prevalence in line with what would be expected. Larger $\beta$ indicates faster spread of the pathogen while larger $\gamma$ values result in lower prevalence in the whole network at steady state. The system wide effect is smaller for relative changes in $\beta$ than in $\gamma$.
		
		The length of the initial infection phase (i.e. the time when the percentage of the colonized patients is below 10\% of the final prevalence) seems to be highly dependent on the parameter $\beta$. This effect is visible in all the presented simulations.
		If we look closer at the effect of varying recovery rates $\gamma$ (Figure \ref{fig:prevelence:gamma:b}), we find a similar behaviour, but the changes are milder.

		\FloatBarrier

	\subsection{Impact of indirect transfers}\label{sec:imp:gamma}

		\begin{figure}
			\centering
			\includegraphics[width=\textwidth]{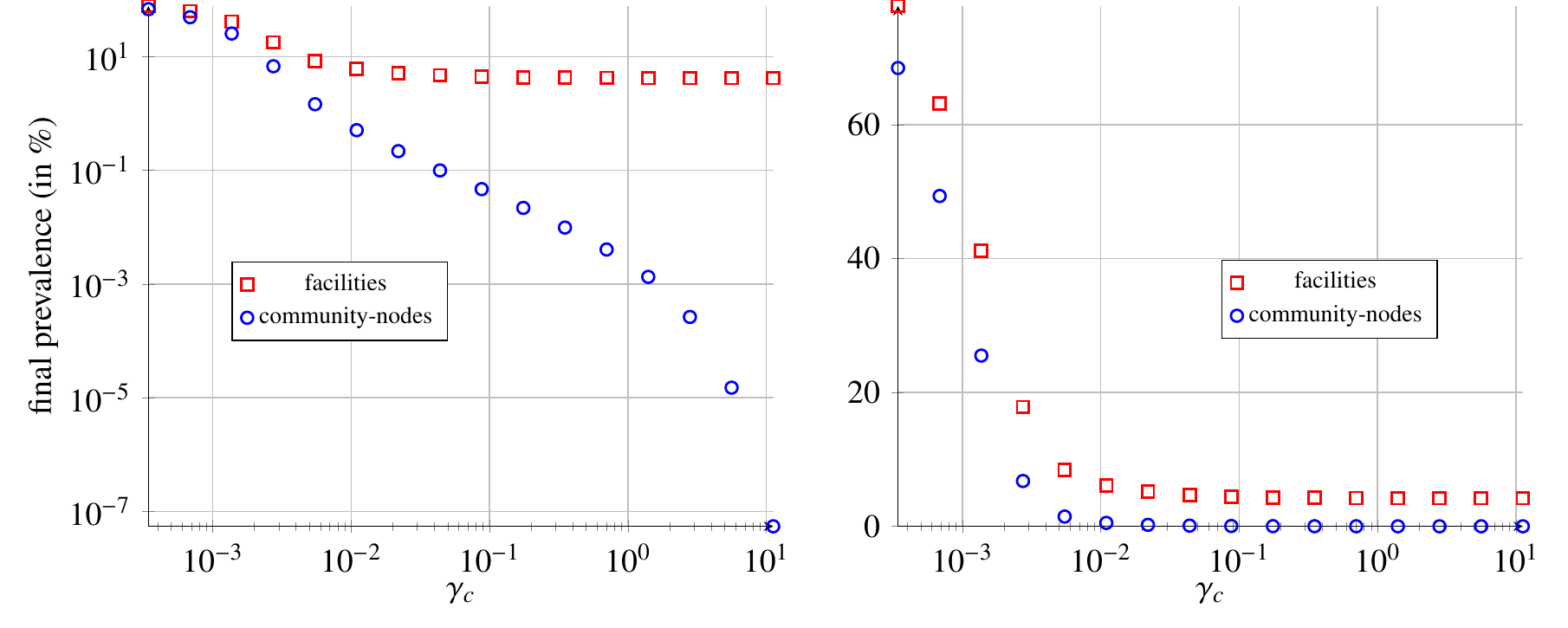}
			\caption{Dependence of network prevalence  in healthcare facilities and corresponding community-nodes (reported for the last day) on different values of $\gamma_c$ parameter in community-nodes, if $\gamma_h=1/365$ for all hospitals.
			 Initial fraction of the colonized people is 0.1\%  uniformly distributed in the whole population.
			(left) Data presented in log-log scale. (right) Data presented in semilog(x) scale.
			}
			\label{fig:gamma:inf}
		\end{figure}

		\begin{figure}
			\centering
			\begin{subfigure}[t]{0.49\textwidth}
				\centering
				\includegraphics[height=5.9cm]{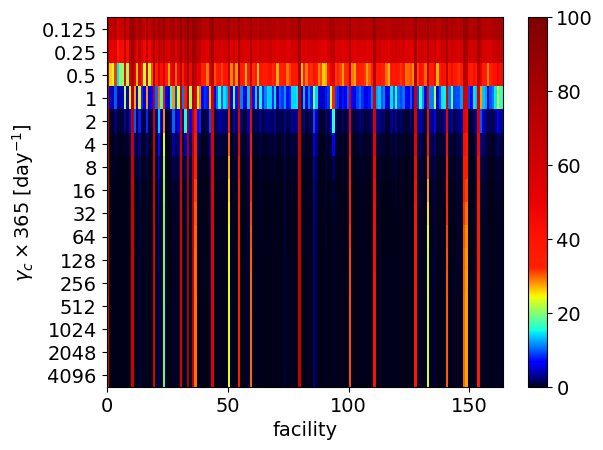}
				\caption{ \label{fig:prevel:gammac:all:a}}
			\end{subfigure}
			\begin{subfigure}[t]{0.49\textwidth}
				\centering
				\includegraphics[height=5.9cm]{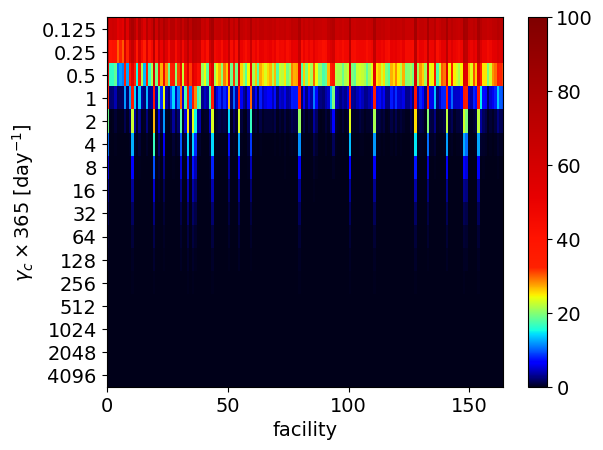}
				\caption{ \label{fig:prevel:gammac:all:b}}
			\end{subfigure}
			\caption{Final network prevalence depending on the value of $\gamma_c$ parameter (vertical axis) for: (a) all considered hospitals, (b) corresponding community nodes. 
				Initial condition for all simulations is the same: 0.1\% uniformly-distributed colonized patients. Facilities are sorted by increasing average size.
				\label{fig:prevel:gammac:all}
			}
		\end{figure}

		In order to assess the role of indirect transfers for pathogen spread, in this section we change the recovery rates in community nodes ($\gamma_c$), which in generally differ from the recovery rates in corresponding hospitals ($\gamma_h$), as indicated in Table~\ref{tab:sis:params} (set type: indirect transfer impact).

		In Figure ~\ref{fig:gamma:inf}, we show how the prevalence is affected. Starting from the prevalences at level of $\approx77\%$ and $\approx68\%$ for the communities and healthcare facilities, respectively, the prevalences decrease monotonically with increasing $\gamma_c$. For the simulations with $\gamma_c>\SI{128/365}{day^{-1}} \approx \SI{0.35}{day^{-1}}$, network prevalence in hospitals, understood as percentage of infectious individuals, stabilizes at the level of about 4.2\%. For community nodes it systematically decreases falling below 0,1\% for $\gamma_c$  greater than $16/365 \;\mathrm{day}^{-1} \approx 4.38\times10^{-2} \;\mathrm{day}^{-1}$.

		As presented in Figure~\ref{fig:prevel:gammac:all}, the final prevalence between the facilities and corresponding communities varies. Moreover, this diversity also changes depending on the value of $\gamma_c$ parameter. Nevertheless, there is a small fraction of facilities, where the prevalence is very high independently of the community recovery rate. Similar effect we observe for communities. However, in this case for sufficiently high $\gamma_c$ all communities have very low prevalence.



		\begin{figure}
			\begin{subfigure}[t]{0.49\textwidth}
				\centering
				\includegraphics[width=1\linewidth]{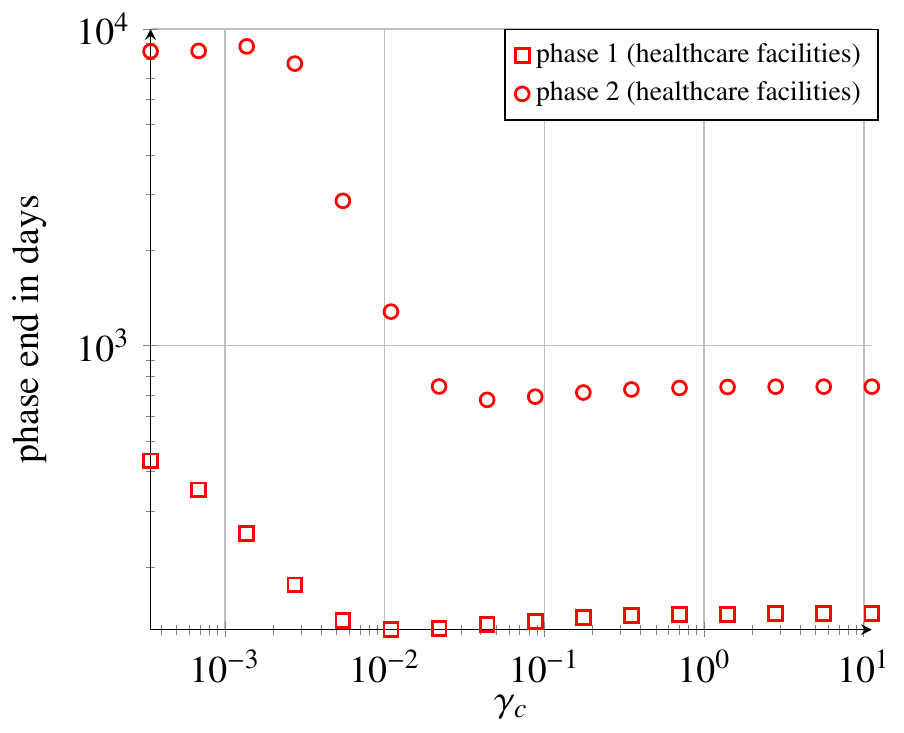}
				\caption{ \label{fig:ghphase-endgammac:a}}
			\end{subfigure}
			~
			\begin{subfigure}[t]{0.49\textwidth}
				\centering
				\includegraphics[height=6.5cm]{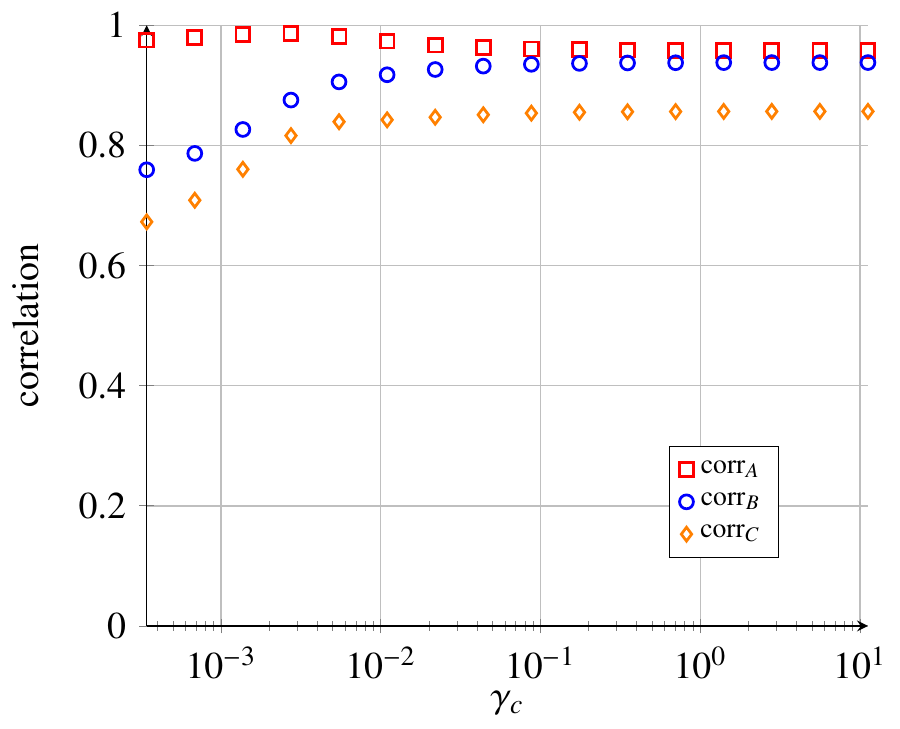}
				\caption{ \label{fig:ghphase-endgammac:c}}
			\end{subfigure}
			\caption{(a) System-wide phase ends in healthcare facilities for different $\gamma_c$. (b) Dependence of the correlation coefficients on $\gamma_c$ value (semi-log scale), 
			corr$_A$ is the correlation coefficient between final hospital prevalences and corresponding communities prevalences; 
			corr$_B$ -- correlation coefficient between final hospital prevalences and 	 average length of stay in hospitals;
			corr$_C$ -- correlation coefficient between final community prevalences and  average length of stay in corresponding hospitals. Initial condition for all simulations is the same 0.1\% uniformly-distributed colonized patients.}
			\label{fig:ghphase-endgammac}
		\end{figure}

		Another feature affected by transmission of the pathogen via indirect transfers is the time needed to reach the final state (see Figure~\ref{fig:ghphase-endgammac:a}). What is interesting, time to reach the end of phase 1 (prevalence lower than 10\% of the final prevalence) shows for small $\gamma_c$ a monotone decrease, while it is not the case for the end of phase 2 (i.e. prevalence < 99.9\% final prevalence).

		Additional study of the dependences of variety of the correlation coefficient for different $\gamma_c$ values manifests that correlation coefficients between: final prevalence in the hospitals and community nodes (corr$_A$), final prevalence in the hospitals and the average length of stay in hospitals (corr$_B$) and final prevalence in communities and the average length of stay in hospitals (corr$_C$), are strongly positive for all considered $\gamma_c$ values, see Figure~\ref{fig:ghphase-endgammac:c}.
		Moreover, again we see accurate functional dependence of the value of community recovery rate. For small $\gamma_c$, strongest correlation is pronounced between final prevalence in the hospitals and corresponding community nodes. However, it slightly decreases with increasing $\gamma_c$, while at the same time correlations between final prevalence in the hospitals (or in communities) and the average length of stay in hospitals increases.
		On the other hand, for sufficiently large community recovery rate ($\gamma_c>0.1 \;\mathrm{day}^{-1}$), when pathogen spread via community nodes is minor or neglectable, all the correlations stabilise at certain levels and become independent of $\gamma_c$. The investigation of correlations between:  final prevalence in hospitals (or community nodes) with in- or out-degree of hospital nodes (or out-degree of community nodes) brought no statistically significant results.

		\FloatBarrier

	\subsection{Influence of the length of stay on the prevalence}\label{sec:av:stays}

\begin{figure}
			\begin{subfigure}[t]{0.49\textwidth}
			\includegraphics[width=\linewidth]{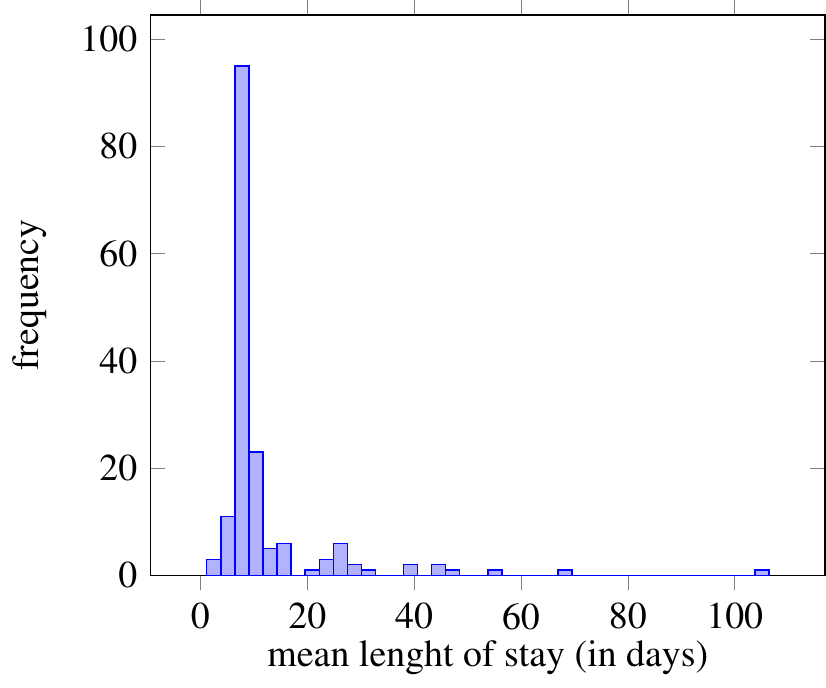}
			\caption{\label{fig:length_all_hosp_histogram:a}}
			\end{subfigure}
		~	
			\begin{subfigure}[t]{0.49\textwidth}
			\includegraphics[width=\linewidth]{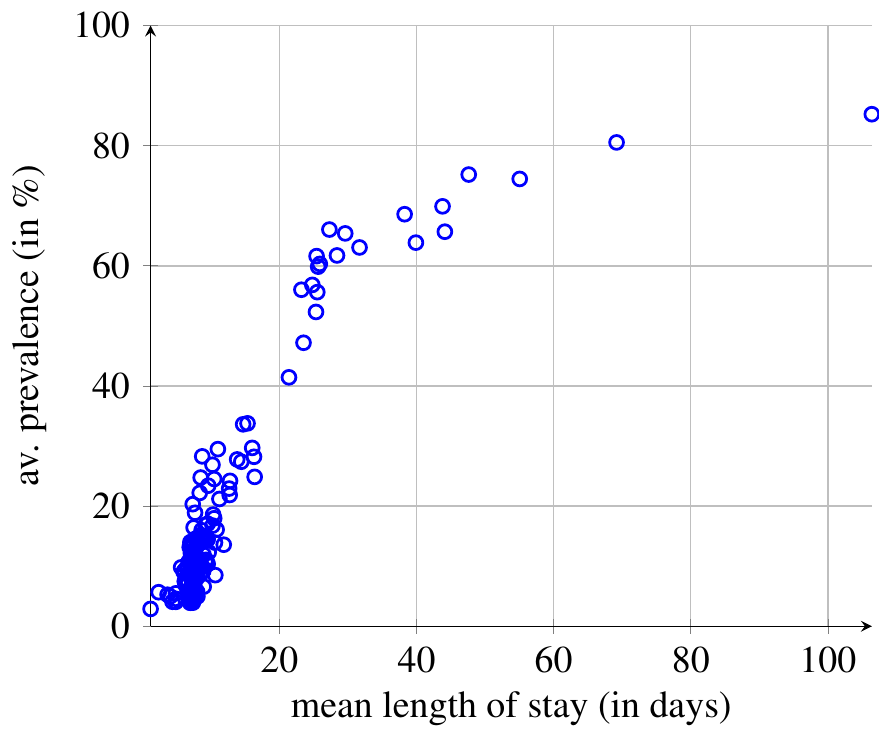}
			\caption{\label{fig:length_vs_prevalence:b}}
			\end{subfigure}	
			\caption{
			(a) Distribution of length of stay (in days) for all considered hospitals (164 units).
			(b) Average length of stay vs. average prevalence calculated for each hospital  (all considered units).\label{fig:mix:length_all_hosp_histogram_length_vs_prevalence}
		}
		\end{figure}

		In Figure~\ref{fig:length_all_hosp_histogram:a} we present the distribution of average length of stays (in days) for considered hospitals, as discussed in Section~\ref{sec:trans}. In most cases, it is between 7--9 days.

		In Figure~\ref{fig:length_vs_prevalence:b} the dependence between the average length of stay (all considered hospitals) and the average prevalences calculated for each hospital are presented. The prevalence for each hospital was calculated based on the simulations presented in Sec.~\ref{sec:inital:infection} by taking the average from the simulations where the origin of the colonization was changed (164 simulations in total).  The obtained dependence is rather non-linear however most points are focused in one area which strongly affecting results of regression methods.

		\FloatBarrier
\section{Conclusion and Discussion}	

	We investigated the dynamics of MRSA as an example for multidrug resistant bacteria in proposed network model of patient traffic in the German healthcare system taking into account indirect patient transfers and patients staying for some time in the community before readmissions to hospitals.
	Due to the characteristic of pathogens, for which transmission occurs mostly in hospitals, patient exchange between facilities is a main driver of their spread.
	We also showed how to derive the graph for a given healthcare system from an anonymized admission/discharge database, as the transfer databases are not available.
	With every healthcare facility we associate two graph nodes, one corresponding to the patients remaining in the facility, and the other for former patients waiting for readmission. 
	To examine the proposed approach, we presented numerical stimulations for a healthcare system of Lower Saxony, Federal Republic of Germany.

	Our analysis of the AOK Lower Saxony anonymized database for years 2008--2015 shows that patient transfers are mostly well balanced and thus we can use deterministic model and in particular Markov chain to describe the patients movement within the network. Nevertheless, derived probability matrix defining Markov process is not symmetric, indicating the directed graphs should be used instead of the indirected one.
	It should be pointed out that a Markov probability matrix can be non-symmetric even for perfectly balanced transfers because of different sizes of facilities, and thus different probability related to the same patient numbers. Although we propose simplified deterministic transfer model, stationary probability distribution (eigenvector corresponding to eigenvalue equal to 1) are close to values estimated directly from data, underlining the correctness of proposed approach.

	We found that only 6\% of patients in our underlying dataset are transferred directly.
	The remaining 94\% spend some time in the community before readmission. 
	From these transfers, we deduced a Markov process transfer probability matrix, which in combination with a SIS intra-hospital model determines the dynamics of the pathogen transmission process within the hospital network. 
	We found that a typical spread pattern in such a network consists of three phases.
	In the first phase (\emph{initial}), the rate of colonized patients is close to 0 and the length of this phase is strongly associated with the contact rate $\beta$.
	In the second phase (\emph{transient}), the prevalence rapidly increases, and then it reaches a stable level, which is the final third phase (\emph{stable}).
	Length of these phases and final prevalence depends on $\beta$ and on the recovery rate $\gamma$. In addition we observed that if we look at the stable-state colonization rate of the individual facilities, it is not uniform (despite uniform $\beta$ and $\gamma$).
	Mean length of stay is one of the determinants of prevalence at steady state (understood as the state after sufficiently long time, when the changes are not pronounced anymore) with the longer stays being associated with higher prevalence.
	The node position in the network may also play an important role. However, our analysis of node degrees shows no statistically significant correlation between in-degrees (or out-degrees) and the prevalence in hospital nodes or corresponding community nodes.
	Additionally, we observe that the final prevalence in units is independent of the facility, from which the pathogen spread was initiated in, however the initial dynamics may slightly differ.

	In the main part of our analysis we found that taking the indirect transfers into account has a significant effect on the final prevalence in the healthcare system.
	We show that restricting the pathogen spread to direct-transfers only, leads to 4.2\% final system wide prevalence, while including indirect transfers leads to a prevalence increase to almost 18\%. This example demonstrates that this additional pathogen transmission channel may lead   to relevant increases in the prevalence and cannot be ignored. Limiting the simulation to the healthcare system, and ignoring the indirect transfers and the community, may result in underestimation of the prevalence. There are also differences in the transmission dynamics. For high $\gamma_c$ (corresponding to transmission of MRSA via the direct transfers only) the system reaches the stable state in 2--3 years, while with indirect transfers it needs about 20 years to reach almost 18\% prevalence. Dependent on the type of pathogen studied and its true clearance rate in the general population, network flows representing indirect transfers within a pre-defined time frame need to be taken into account to understand the true transmission dynamics in a healthcare network, and the effectiveness of potential infection control measures.

	Moreover, using proposed model, we have also found a non-linear concave functional dependence of the prevalence in the hospitals on the average length of stay in that facilities, which can be an useful tool for estimation of the prevalence in given hospitals, knowing the average length of patient stays of the considered facility. However, this dependence is based strongly on the network structure, and in the future we will address that issue studying different hospital networks.

	One example of emerging multi-resistant pathogens, which are more complex to model than MRSA, are multi-resistant gram-negative bacteria, especially Enterobacteriacaeae. In \cite{Gurieva2017}, the prospective surveillance data from 13 European intensive care units were used to estimate transmission rates of Escherichia coli and E. coli Enterobacteriaceae. This group of pathogens share resistance genes, but have very different transmission characteristics, transmission ways and clearance rates, making the correct choice of time frames for indirect transfers more difficult~\cite{Gurieva2017}.

	Infection control by adjustment of the network structure (as e.g. proposed by Donker~et~al. in~\cite{Donker2012}), provide another level of complexity which can be considered in combination with classic infection control measures focusing on individual healthcare facilities. In~\cite{Lee2011}, Lee~et~al. have shown that distribution of infection control measures based on network characteristics can improve effectiveness of infection control measures and save resources at the same time. On the other hand, our analysis showed that the decrease of the transferability of the pathogen via the community does not significantly influence the correlation between the final network prevalence (in the facilities and corresponding communities) and average lengths of stays in hospitals. 

	The study presented in this paper is performed for a single federal state in Germany. Provided that there are available data, the same may be done for additional regions, group of countries. One further challenge will be to determine patient transfer between administrative regions. Our model can be further expanded by taking into account infection control measures: this require more detailed modelling at the intra-hospital level. Screening, isolation  of colonized patients and introduction of effective treatment  may be simulated for example by decreasing $\beta$.
	Also, additional stratification of risk may be added by similar means.


\section{Appendix}\label{sec:app}

	\subsection{Lemma \ref{lemma:regular:stochastic}}
	
	The proof of the following lemma can be find in~\cite{Lonc2019}.
	\begin{lemma}
	\label{lemma:regular:stochastic}
	Assume that $A=[A_{i j}]_{i,j=1}^{k}$ is a $k \times k$ dimensional real matrix such that
	\begin{enumerate}
		\item $\forall i,j \in \{1, \ldots, k\}$, $i \neq j$,  $A_{i j} \geq 0$,
		\item $\forall j \in \{1, \ldots, k\}$  $A_{j j} > 0$,
		\item $\forall i \in \{1, \ldots, k\}$  $\sum_{j=1}^k A_{i j} = 1$,
		\item $\forall i,j \in \{1, \ldots, k\}$ $\exists \,i_0 = i, i_1, \ldots, i_{n-1}, i_n = j$ such that $\forall m \in \{1, \ldots, n\} \quad A_{i_{m-1} i_m} > 0$.
	\end{enumerate}
	Then $A$ is a stochastic regular matrix.
	\end{lemma}
	This lemma may be applied to hospital transfer probabililty matrices of this paper. 
	Clearly, Assumption 1 of Lemma~\ref{lemma:regular:stochastic} is satisfied due definition (\ref{eq:aij:obl}), as all the elements are non-negative.
	Assumption 2 is more subtle, as in general $A_{jj}$ can be equal to $0$.  It is however rather unlikely, as due to  (\ref{eq:aij:obl}) it would imply that no patients would stay in the healthcare facility overnight ever. Assumption 3 is a direct consequence of definition~(\ref{eq:A:diag}). Assumption 4 means that for every two facilities, there is some (potential) transfer path between them. Actually, it is not necessary that any patient follow this whole path, but there must be some patient transfer for every component. Thus, the transfer path must exist between every two facilities, in both directions.

	\subsection{Transfer classification algorithm}
		\label{sec:transfer:algorithm}

		To classify  patient transfers properly, we propose a heuristic algorithm, which detects and classifies transfers from the available claims database in a general manner.
		Please note that we use in description of the algorithm: $i$ to denote days, $l$ for facilities, $p$ for patients and symbols $A_i$, $d$, $D$, $H_i$, $L_i$, $P_i$  defined in the text. This notation is local, i.e. it will be used in description of this algorithm , however some of these symbols will be later used for different notions.

		Our algorithm operates at the basis of individual patients and works as follows. 
		Let $H$ denote a set of healthcare facilities.
		Let us take a patient and the admission/discharge table related to their hospital stays. Then by $D := \{0, 1, \ldots, d\}$  we denote a set of days at which this patient stays in any healthcare facility. We assume that the order of the indices $D$ corresponds to the order of days, but it is possible that these days are not consecutive, i.e. days $i$ and $i+1$ may actually be separated by some period of time.

		Then let $H_i \subset H$ be a set of healthcare facilities in which the patient formally stays at day $i$, i.e. there is at least one entry in the considered admission/discharge dataset, which corresponds to this patient while the day $i$ is between the admission date and discharge date of this entry (both inclusive). Moreover, we define $H_{-1} := \emptyset$ and a set of newly-arrived facilities as $A_i := H_i \backslash H_{i-1}$ while a set of recently-left facilities is denoted by $L_i :=  H_{i-1} \backslash H_i$.

		$H_i$ can have than one element when entries overlap and patients are formally assigned to multiple facilities at the same day. To track such situations, we also define sets $P_i \subset H$, which correspond to facilities in which we think the patient is present at day $i$. These sets will be determined by the proposed heuristic.

		If $i-1$ and $i$ are consecutive days, we check every new facility $a \in A_i$. Then for every $p \in P_{i-1}$ we deduce a transfer from $p$ to $a$.

		Then for any facility $l \in L_i \cap P_{i-1}$ which the patient formally leaves at day $i$ and in which they were present at day $i-1$, and for any facility $h \in H_i$ in which they formally are at day $i$, we deduce a transfer from $l$ to $h$. We take $h \in H_i$ instead of for example $h \in P_{i-1}$.
		To account for the case, when a patient has a temporary transfer to another facility. When the new stay finishes, patients return to the original facility, in which they are now considered as not present (the facility is not in $P_{i-1}$), but formally they have a bed reserved there (it is in $H_i$).

		We determine $P_i$ as follows:
		\begin{itemize}
			\item If there is no new admission or discharge ($A_i = L_i = \emptyset$), then nothing changes ($P_i := P_{i-1}$).
			\item If there are new admissions ($A_i \neq \emptyset, L_i = \emptyset$), then we deduce that the patient is transferred to these new locations and $P_i := A_i$. We cannot determine without additional information whether they are still present in other places from $H_i$ or not, so we assume that beds are only held for them.
			\item If there are discharges ($L_i \neq \emptyset$), then any facility left is removed from $P_i$; all current facilities are added to this set, as we assume that the patients are returning to the original facilities. So, depending on whether there were new admissions ($A_i \neq \emptyset$) or not ($A_i = \emptyset$), we define $P_i = (A_i \backslash L_i) \cup H_i$ or $P_i = (P_{i-1} \backslash L_i) \cup H_i$.
		\end{itemize}

		On the other hand, if $i$ and $i+1$ are not consecutive days, i.e if there is at least one day in between, then for every $p \in P_{i-1}$ and for every $a \in A_i$ we deduce a transfer from $p$ to $a$ and we set $P_i := A_i$. Note that it is possible that $P_{i-1} \cap A_i \neq \emptyset$, thus we can deduce indirect auto-transfers also. Since we need some initial value, we define $P_{-1} := \emptyset$ and we assume that $-1$ and $0$ are not consecutive days.

		This procedure is executed for each patient from the dataset; afterwards transfers are accumulated.

\section{Acknowledgements} 

	This publication was made possible by grants from following national funding agencies: National Science Centre, Poland, Unisono: 2016/22/Z/ST1/00690 (University of Warsaw, Faculty of Mathematics, Informatics and Mechanics, Institute of Applied Mathematics and Mechanics) and 01KI1704C (Martin-Luther-University Halle-Wittenberg, Medical Faculty, Institute of medical epidemiology, biostatistics and informatics) and the Netherlands ZonMw grant number 547001005 (Julius Centre, University Medical Centre Utrecht) within the 3rd JPI ARM framework (Joint Programming Initiative on Antimicrobial Resistance) cofound grant no 681055 for the consortium EMerGE-Net (Effectiveness of infection control strategies against intra- and inter-hospital transmission of MultidruG-resistant Enterobacteriaceae).

	We thank the AOK Lower Saxony for providing anonymized record Data.

\bibliographystyle{siam}
\bibliography{literature}

\end{document}